\documentclass[sigconf]{acmart}
\AtBeginDocument{%
  }

\copyrightyear{2025}
\acmYear{2025}
\setcopyright{cc}
\setcctype{by-nc-nd}
\acmConference[CHI '25]{CHI Conference on Human Factors in Computing Systems}{April 26-May 1, 2025}{Yokohama, Japan}
\acmBooktitle{CHI Conference on Human Factors in Computing Systems (CHI '25), April 26-May 1, 2025, Yokohama, Japan}\acmDOI{10.1145/3706598.3713395}
\acmISBN{979-8-4007-1394-1/25/04}

\newcommand{\tool}{TeamVision}
\usepackage{tabularray}
\usepackage{graphicx}
\usepackage{hyperref}
\usepackage{textcomp}
\usepackage{pifont} 
\usepackage{xcolor} 
\usepackage{fancybox}

\newcommand{\filledcircleone}{{\scalebox{1.5}
{\textcolor{lightgray}{\raisebox{-0.2em}{\ding{108}}}}
\hspace{-1.35em}
\textcolor{black}{\textbf{\tt 1 }}}}

\newcommand{\filledcircletwo}{{\scalebox{1.5}
{\textcolor{lightgray}{\raisebox{-0.2em}{\ding{108}}}}
\hspace{-1.35em}
\textcolor{black}{\textbf{\tt 2 }}}}

\newcommand{\filledcirclethree}{{\scalebox{1.5}
{\textcolor{lightgray}{\raisebox{-0.2em}{\ding{108}}}}
\hspace{-1.35em}
\textcolor{black}{\textbf{\tt 3 }}}}

\newcommand{\customovalbox}[1]{%
    \setlength{\fboxsep}{2pt}
    \colorbox{white}{\textcolor{black}{\ovalbox{#1}}}%
}

\newcommand{\allphases}{\customovalbox{\small \tt all}}
\newcommand{\phone}{\filledcircleone~}
\newcommand{\phtwo}{\filledcircletwo~}
\newcommand{\phthree}{\filledcirclethree~}

\newcommand{\filledcircleA}{{\scalebox{1.5}{\raisebox{-0.2em}{\ding{108}}}\hspace{-0.85em}\textcolor{white}{\textbf{\tt A }}}}

\newcommand{\filledcircleB}{{\scalebox{1.5}{\raisebox{-0.2em}{\ding{108}}}\hspace{-0.85em}\textcolor{white}{\textbf{\tt B }}}}

\newcommand{\filledcircleC}{{\scalebox{1.5}{\raisebox{-0.2em}{\ding{108}}}\hspace{-0.85em}\textcolor{white}{\textbf{\tt C }}}}

\newcommand{\itemA}{\filledcircleA}
\newcommand{\itemB}{\filledcircleB}
\newcommand{\itemC}{\filledcircleC}

\newcommand{\barchart}{priority chart}
\newcommand{\wardmap}{ward map}
\newcommand{\SNA}{sociogram}
\newcommand{\EN}{communication network}
\begin{document}

\title[TeamVision: An AI-powered Learning Analytics System for Supporting Reflection]{\tool: An AI-powered Learning Analytics System for Supporting Reflection in Team-based Healthcare Simulation}

\author{Vanessa Echeverria}
\email{vanessa.echeverria@monash.edu}
\orcid{0000-0002-2022-9588}
\affiliation{%
  \institution{Monash University}
  \city{Melbourne}
  \state{VIC}
  \country{Australia}
}
\affiliation{%
  \institution{Escuela Superior Politécnica del Litoral}
  \city{Guayaquil}
  \country{Ecuador}
}

\author{Linxuan Zhao}
\affiliation{%
  \institution{Monash University}
  \city{Melbourne}
  \country{Australia}
}

\author{Riordan Alfredo}
\affiliation{%
  \institution{Monash University}
  \city{Melbourne}
  \country{Australia}
}

\author{Mikaela Milesi}
\affiliation{%
  \institution{Monash University}
  \city{Melbourne}
  \country{Australia}
}

\author{Yueqiao Jin}
\affiliation{%
  \institution{Monash University}
  \city{Melbourne}
  \country{Australia}
}

\author{Sophie Abel}
\affiliation{%
  \institution{Macquarie University}
  \city{Sydney}
  \country{Australia}
}

\author{Jie Fan}
\affiliation{%
  \institution{Monash University}
  \city{Melbourne}
  \country{Australia}
}

\author{Lixiang Yan}
\affiliation{%
  \institution{Monash University}
  \city{Melbourne}
  \country{Australia}
}

\author{Xinyu Li}
\affiliation{%
  \institution{Monash University}
  \city{Melbourne}
  \country{Australia}
}

\author{Samantha Dix}
\affiliation{%
  \institution{Monash University}
  \city{Melbourne}
  \country{Australia}
}

\author{Rosie Wotherspoon}
\affiliation{%
  \institution{Monash University}
  \city{Melbourne}
  \country{Australia}
}

\author{Hollie Jaggard}
\affiliation{%
  \institution{Monash University}
  \city{Melbourne}
  \country{Australia}
}

\author{Abra Osborne}
\affiliation{%
  \institution{Monash University}
  \city{Melbourne}
  \country{Australia}
}

\author{Simon Buckingham Shum}
\affiliation{%
  \institution{University of Technology Sydney}
    \city{Sydney}
  \country{Australia}
}

\author{Dragan Gašević}
\affiliation{%
  \institution{Monash University}
  \city{Melbourne}
  \country{Australia}
}

\author{Roberto Martinez-Maldonado}
\email{roberto.martinez-maldonado@monash.edu}
\affiliation{%
  \institution{Monash University}
  \city{Melbourne}
  \state{VIC}
  \country{Australia}
}

\renewcommand{\shortauthors}{Echeverria et al.}

\begin{abstract}
 Healthcare simulations help learners develop teamwork and clinical skills in a risk-free setting, promoting reflection on real-world practices through structured debriefs. However, despite video's potential, it is hard to use, leaving a gap in providing concise, data-driven summaries for supporting effective debriefing.
Addressing this, we present TeamVision, an AI-powered multimodal learning analytics (MMLA) system that captures voice presence, automated transcriptions, body rotation, and positioning data, offering educators a dashboard to guide debriefs immediately after simulations.
We conducted an in-the-wild study with 56 teams (221 students) and recorded debriefs led by six teachers using TeamVision. Follow-up interviews with 15 students and five teachers explored perceptions of its usefulness, accuracy, and trustworthiness. This paper examines: i) how TeamVision was used in debriefing, ii) what educators found valuable/challenging, and iii) perceptions of its effectiveness. Results suggest TeamVision enables flexible debriefing and highlights the challenges and implications of using AI-powered systems in healthcare simulation.
\end{abstract}




\maketitle

\begin{figure*}[htp!]
  \includegraphics[width=\textwidth]{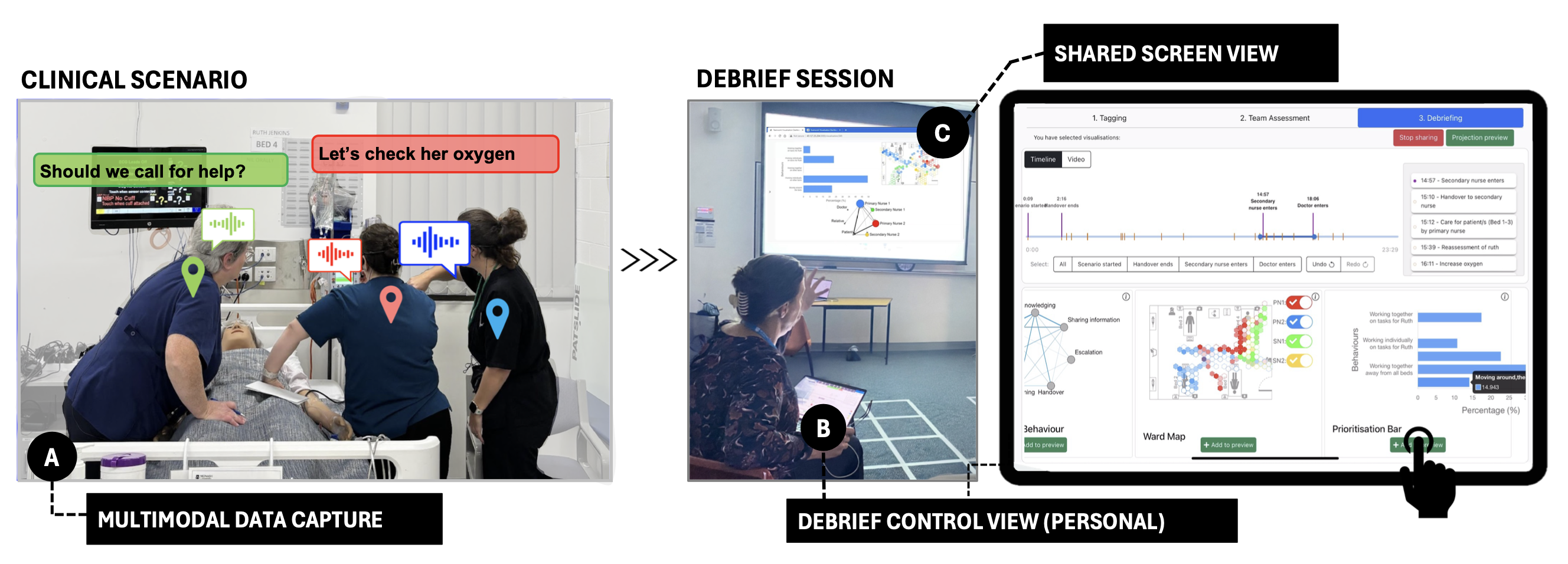}
\caption{\tool: An AI-powered multimodal system that captures (A) human observations, real-time positioning, and audio data from nursing students during high-fidelity clinical simulations. It provides an analytics dashboard. (B) Educators can filter and customise data visualisations to fit their approach, while (C) students view performance analytics on a shared screen during the debrief.}
\Description{Top: three nursing students attending a simulated patient. Right: a teacher using the analytics during the debrief.}
  \label{fig:teaser}
\Description{Figure A. Healthcare professionals in a clinical scenario with a patient on a bed, followed by a debrief session using a shared screen with graphs and data. 
Figure B: The image shows a tablet with a doctor and patient, illustrating AI analytics for nursing students during clinical debriefing.
}
\end{figure*}
\section{Introduction}
Learning to work effectively in teams is particularly crucial in professions like firefighting \cite{Dawes_Cresswell_Cahan_2004}, healthcare \cite{Mazzocco_2009}, and strategic training \cite{salas2018science}, where communication and physical coordination are essential for ensuring safety and minimising errors \cite{salas2018science, Morey_2002}. In these contexts, teamwork is inherently complex, involving cognitive, affective, verbal, and behavioural interactions among team members, often supported by various technologies and tools \cite{Kozlowski_2018}. As a result, developing effective teamwork skills has been a central focus of CSCW and HCI research for many years \cite{carroll2005teamwork,Fitzpatrick2013cscw}.


In the context of healthcare education, simulation-based learning has proven to be an effective strategy, allowing medical and nursing students, as well as professionals, to practice clinical and teamwork skills in a safe, realistic environment \cite{murphy2016}. Using interactive systems such as specialised high-fidelity manikins, software, or standardised human patients, and often involving actors, this approach enables healthcare procedure practice with minimal risk, both in-situ in clinical settings \cite{Truchote059442} and in dedicated simulation rooms \cite{Moslehi_2022}. These team-based simulations incorporate reflective practices that help learners develop technical, teamwork, and communication skills \cite{Morey_2002,murphy2016}.
Simulations are typically followed by a reflective debrief, led by a trained facilitator or educator, where participants discuss their performance. These sessions engage learners in reflective practice, encouraging critical analysis of their actions, decisions, and outcomes, as outlined in reflection theory \cite{schon2017reflective}. The educator guides the discussion with questions about performance, emotional responses, successes, and areas for improvement \cite{salas2018science}.

Yet, debrief educators often rely solely on their memories and observations, which can result in the omission of crucial technical skills and team dynamics, such as nurse-doctor or nurse-patient communication, essential for positive patient outcomes \cite{Morey_2002}. To address this, video recordings with annotations have been introduced to support evidence-based reflective practices, enabling participants to revisit critical events from the simulation \cite{hall2017best,Abeer2018}. While these approaches can arguably \cite{ROSSIGNOL2017145,Cheng2014debriefing,Abeer2018,KROGH2015180} enhance reflective debriefing for some educational simulations, they present challenges: video analysis is time-consuming \cite{sablic2021video}, may introduce distractions, and can be costly \cite{schertzer2023videoDebriefing}. Additionally, the educator's subjective perspective may still dominate \cite{fraser2018cognitive}, and the effectiveness of video-based debriefs depends on the educator’s skills in annotating and using video replays effectively \cite{schertzer2023videoDebriefing}. These challenges limit deeper reflection opportunities \cite{hall2017best}, and the complexity and resource demands of video tools have hindered their widespread use \cite{rosvig2023,KROGH2015180}. Recent reviews consistently underscore the need for more evidence on teamwork in healthcare simulation, quantitative performance measures, and scalable, standardised debriefing tools to better support debriefs \cite{Moslehi_2022, cant2017valueOfSim, hegland2017simSLR,hildreth2023telEmergency, Martin_2020}.

Advancements in sensing technologies, multimodal learning analytics, and artificial intelligence (AI) offer new opportunities for improving data capture in learning spaces \cite{DiMitri2024bjet,giannakos2022sensor,Giannakosrole2023}, including generating evidence on teamwork in healthcare simulations \cite{Martinez-Maldonado2017analyticsmeet,ochoa_multimodal_2022}, and addressing limitations of traditional video replay, observation, and annotation methods. These technologies can extract salient aspects of collaboration, such as visualising learner socio-spatial behaviours \cite{yan2023sena} or identifying workflow bottlenecks in team-based emergency simulations \cite{Petrosoniak2018}. AI algorithms can analyse team dynamics, detecting non-verbal behaviours in surgical training \cite{Harari_2024} or using speech-to-text and large language models (LLMs) to identify communication practices like closed-loop communication \cite{Jagannath2022,Zhao2024BJET}. These AI-powered analytics can be presented via user-friendly interfaces, such as learning analytics dashboards \cite{kaliisa2023,Sun2019mydata}, to augment \cite{Khakurel2022augmenting} educator observations and support reflective debriefing. However, this vision remains to be fully realised.

Inroads have been made in integrating AI and analytics into simulation-based learning to study and support teamwork in healthcare, though most of this work has focused on advancing educational knowledge and theories rather than creating interfaces for use in debriefs. For example, some research has modelled how nurses move in space \cite{fernandez2021can,echeverria2018where}, analysed voice detection patterns in high- vs low-performing teams \cite{zhao2024lak}, examined nurse team dialogue \cite{Zhao2024BJET}, and tracked where nurses look during emergency simulations \cite{Davalos2323}. 
Recent efforts have started to explore the design of data interfaces to help learners reflect on errors \cite{echeverria2019collaboration}, positioning dynamics \cite{fernandez2022classroom}, and teamwork behaviours \cite{milesi2024comics}, but these studies were conducted in controlled lab environments, far removed from actual debriefing situations, and focused on personal reflection without an educator facilitating the debrief. Nevertheless, they demonstrated the potential benefits of data to support reflection after healthcare simulations.
The closest work towards analytics debriefing integration is the work by \cite{Echeverria2024LAK}, which developed a prototype with three static visualisations shown during the debrief. While results suggested using data was promising, the static visualisations were not always relevant. The system needed to adapt to existing debriefing practices, enabling data navigation through annotations and allowing educators to filter and select relevant visualisations. These works highlight the opportunity and need to design AI-powered analytics interfaces that align with current debriefing practices to augment reflective debriefing.

This paper addresses this gap in the literature by presenting \tool, an AI-powered system that captures sensor data (voice presence, automatically coded transcriptions, body rotation, and positioning) and human annotations. \tool~ provides educators with a dashboard offering on-demand visualisations on team prioritisation, teamwork, and communication strategies to guide debriefs after simulations. Five educators participated in a five-month human-centred design process to ensure \tool~ aligned with current reflective and debriefing practices.
To validate \tool~, we conducted an in-the-wild study in an undergraduate nursing course, where educators and students used \tool~ as part of the regular curriculum. We collected multimodal data and real-time annotations from 56 teams (221 students), along with audio recordings and interaction data from six educators during debriefs (n=26). Five of these educators were interviewed to discuss their experience with \tool, focusing on usability, challenges, and its potential to support reflection in healthcare simulation. Additionally, 15 students participated in follow-up interviews to share their views on \tool's usefulness, accuracy, and trustworthiness in supporting reflection.
The analysis examined i) how \tool~ was used by educators during actual debriefs, ii) what educators found valuable or challenging when using \tool~ to support their teaching practice, and iii) both educators' and students' perceptions of \tool’s usefulness, accuracy and trust on the information provided in the visualisations. 
Our findings contribute to HCI literature by demonstrating how AI-powered systems hold the potential to be integrated into real-world educational settings, in particular, to augment and support evidence-based reflective practices in healthcare simulations, offering actionable insights into team dynamics and communication.
\section{Background and Related Works}
\subsection{\textcolor{black}{Supporting} Reflection in Healthcare \textcolor{black}{Simulation}: Benefits and Challenges} 

Debriefing is where much of the learning happens in simulation-based training \cite{fanning2007role,Tannenbaum2013}. The debriefing process allows participants to reflect on their actions, discuss mistakes, and \textcolor{black}{consolidate} learning through guided reflection and feedback, enhancing understanding and application of the skills practised during the simulation \cite{fanning2007role,Tannenbaum2013}, and enabling formative assessment of both individual and team performance \cite{tuticci2018,levett2014systematic}. This aligns with Schon's Reflection Theory \cite{schon2017reflective}, where reflective practice is central to learning from experience, encouraging learners to critically examine their actions and decisions for deeper learning and improvement. Similarly, Dewey's experiential learning framework \cite{dewey1933howwethink} highlights the importance of reflection as a key phase in transforming experience into meaningful knowledge. Dewey emphasised that learning occurs through a cycle of action and reflection, making the debriefing process essential for deriving lessons from the simulation and applying them to future practice.

Yet, conducting team nursing simulations and facilitating guided discussions during debriefs present challenges that may limit opportunities for effective learning and reflection. During the clinical scenario, educators must manage multiple tasks, such as observing team dynamics, comparing the team’s demonstrated behaviours to expected outcomes based on learning objectives (e.g., communication, patient management, task prioritisation), and sometimes operating simulation technology, leading to increased mental demands and divided attention \cite{fraser2018cognitive,Tannenbaum2013}. Additionally, educators may find it difficult to capture all relevant behaviours for discussion in debriefs, particularly in complex scenarios with multiple participants, as these mental demands can quickly escalate \cite{fraser2018cognitive}. Furthermore, educators often have limited time to prepare for debriefing sessions, relying on their memory and observations \cite{fraser2018cognitive,fanning2007role}.

\textcolor{black}{Various tools have been introduced to support debriefing, providing educators and learners with evidence to support reflection practices.} For instance, video-assisted debriefing tools allow students to review their actions, enhancing self-awareness and fostering reflection on non-technical skills such as communication and teamwork \cite{MACLEAN201915,REED2013e585}, boosting confidence, and providing a less subjective source for reflection \cite{Abeer2018}. 
\textcolor{black}{Building on these capabilities, current solutions such as video annotation tools (e.g., IRIS Connect\footnote{\url{https://www.irisconnect.com/uk/adaptive-pd/}} and Sibme\footnote{\url{https://www.sibme.com/}}) add a structural approach to review performance, allowing educators and learners to select specific moments from their practice for a more deliberate discussion and reflection, reducing reliance on memory \cite{rich2009video}.} 
\textcolor{black}{Despite their benefits, the effectiveness of these video-assisted tools is highly dependent on several factors. Educators must invest time and develop specific skills, such as selecting relevant footage for review \cite{fanning2007role,Abeer2018,schertzer2023videoDebriefing} and learn how to effectively use the annotation features \cite{Bouten_Haerens_Doren_Compernolle_Cocker_2023}.}
Over-reliance on video replays may also negatively impact learner satisfaction; watching themselves can induce stress, and poorly chosen video clips may divert attention from critical issues \cite{fanning2007role,Abeer2018}.
\textcolor{black}{Partly because of these challenges, video-based debriefing tools are often underutilised, as their complexity and significant resource demands can hinder effective implementation \cite{rosvig2023,KROGH2015180}. This suggests a growing need for scalable and comprehensive data and analytics to provide actionable insights that support educators in facilitating more effective and evidence-based debriefing sessions and ultimately enhancing the learning experience.}

\subsection{Integrating AI and Analytics for Team-based Healthcare Simulation} 

\textcolor{black}{Advancements in sensing technologies, multimodal learning analytics, and AI offer new opportunities for augmenting teamwork and learning within physical learning environments \cite{Khakurel2022augmenting,giannakos2022sensor}.
These technologies can enhance the detection of actions and interactions by tracking individuals, objects, and events \cite{Rogers2006,Dourish_2001,DiMitri2024bjet,Giannakosrole2023}, enabling the development of AI-powered analytics to scrutinise complex learning behaviours and interactions that were previously difficult to capture \cite{giannakos2022sensor,blikstein2013mmla}. By leveraging video, audio, and sensor data, these systems provide scalable support for teaching and learning \cite{ochoa2020controlled,Ogan19}. 
These AI-powered systems have particularly enriched our understanding of team dynamics in educational contexts by investigating verbal and non-verbal communication (e.g., \cite{Pugh2022LAK,stewart2023AIED}), as well as spatial patterns (e.g., \cite{Rosenbaum_2024}).} 
Some progress has been made in integrating AI and analytics to enhance the understanding of teamwork behaviours in healthcare scenarios. 
For example, a recent study demonstrated how effective communication impacts patient outcomes by automatically identifying key speech-related events during trauma resuscitation \cite{Jagannath2022}. Other research examined non-verbal communication indicators such as speaking time and speech overlap, offering insights into team interactions during simulations \cite{zhao2022modelling}. Similarly, automated transcription and coding of team communication provided a comprehensive view of strategies such as information sharing, acknowledging, and questioning \cite{Zhao2024BJET}. These AI models have proven useful for analysing communication patterns and ensuring accurate task allocation in time-sensitive medical scenarios. 

For the case of spatial \textcolor{black}{patterns}, studies have used pose estimation models to assess CPR quality by tracking arm angles and distances between team members, revealing key aspects of proximity and coordination \cite{Weiss2023}. Another study employed manual video tracking to generate heatmaps and movement indicators, highlighting inefficiencies in team coordination that were not immediately visible through verbal communication \cite{Petrosoniak2018}. Additionally, \citet{fernandez2021can} used position data to uncover spatial constructs such as co-presence and socio-spatial formations, visualising these dynamics as ego networks that revealed how nursing students positioned themselves relative to patients and team members.
While the above studies are valuable for identifying \textcolor{black}{team behaviours and performance}, the AI and analytics outputs were not specifically designed to directly support reflective debriefing practices for learners. 

Recent work has begun exploring how AI and analytics can be leveraged to support healthcare \textcolor{black}{education} by making outputs accessible to educators and learners through learning analytics dashboards \cite{kaliisa2023,Sun2019mydata}. For instance, \citet{fernandez2022classroom} designed and evaluated visualisations displaying the team's location, trajectory, and body orientation to help educators and learners better understand spatial interactions. Similarly, \citet{echeverria2019collaboration} developed technological probes that were evaluated by \textcolor{black}{educators} as potential proxies for communication, movement, and task-related events, offering a holistic view of team interactions.

Although these recent studies explored the potential of displaying visualisations to learners and educators, they remain in early design stages and have been tested in controlled environments rather than "in-the-wild" settings \cite{fernandez2022classroom,echeverria2019collaboration}. This gap highlights the need for further research to refine high-fidelity prototypes into mature systems that can be deployed in real-world contexts \cite{Crabtree2013innovationinthewild,Rogers2007hassle,Chamberlain2012inthewild}. Moreover, these studies also lack the flexibility and adaptability to navigate and interact with the data, which is needed for meaningful reflection. The only exception is the work by \cite{Echeverria2024LAK}, which offers very limited interaction and navigation. Their dashboard presents three static visualisations in a slider, allowing educators to switch between them to support reflection during the debrief. Our work aims to respond to educators' need for more flexible, adaptable solutions that can display the temporal progression of team dynamics.

\textcolor{black}{In parallel, commercial video-based solutions (e.g., IRIS Connect, Sibme) have started leveraging AI and analytics to analyse classroom interactions and enhance educators' reflections on their instructional practices. These solutions automate processes such as video clipping and transcription while providing AI-generated insights that identify teaching moments of interest, analyse communication content (e.g., questioning, scaffolding), and generate actionable feedback. Similarly, other commercial training platforms (e.g., SimConverse\footnote{\url{https://www.simconverse.com/}}) have started integrating AI technologies to simulate patient-nurse communication in chat-based scenarios, offering learners a safe and controlled environment to practice and refine their communication skills. However, the effectiveness of these commercial solutions in fostering reflection remains underexplored, particularly in team-based and healthcare education contexts where interactions are more dynamic and complex.}

\subsection{Research Gaps and Contribution to the HCI community} 
Incorporating AI-powered systems into the classroom presents practical challenges that can only be revealed through their use in authentic learning environments \cite{Crabtree2013innovationinthewild,Rogers2007hassle,Chamberlain2012inthewild}. By focusing on AI-powered analytics in real-world educational contexts, this research seeks to move beyond high-fidelity prototypes and explore their practical integration and impact \textit{in-the-wild}. Therefore, this work explores the following question:
\textbf{RQ1:} \textbf{How can {\tool} support educators during in-the-wild simulation debriefings to facilitate tailored reflective discussions with learners?}

Furthermore, although AI-powered systems hold great potential to enhance reflective practices, educators may encounter challenges in adopting and effectively utilising these systems \cite{yan2022scalability}. HCI research has highlighted the inherent complexities involved in designing and implementing AI-powered technologies \cite{Bly1999,yangReexaminingWhetherWhy2020}. Educators face unique challenges when integrating non-traditional technologies into the classroom, which often become fully apparent only after in-the-wild classroom use \cite{martinezmaldonado2023inthewild}.
We explore educators' perceived challenges, as well as the potential perceived value of using AI-powered analytics in the debriefs through our second question: 
\textbf{RQ2:} \textbf{What are educators' perceived benefits and challenges to adoption after using {\tool} in facilitating debriefs?}

Moreover, HCI research has emphasised that trust and usability are critical for the successful adoption of AI-powered systems in educational settings \cite{Shneiderman_2020}. These systems must be carefully tailored to the specific context \cite{Bly1999,yangReexaminingWhetherWhy2020,Liao2023}. A human-centred AI (HC-AI) design approach is essential for navigating AI complexities, including usability, trust, and accuracy of the information presented \cite{Shneiderman_2020}. For these systems to be widely adopted, educators and learners must have confidence that the insights they generate are accurate, actionable, and reliable. There is a need for ecological studies to understand how educators interpret and use this information and how students perceive the usefulness of such information. 
Therefore, we aim to address the following question tailored to the purpose of augmenting simulation debriefs with AI and analytics:
\textbf{RQ3:} \textbf{What are educators' and students' perceptions of {\tool}'s usefulness, accuracy and trust after use in their debriefs?}

\section{Research Context} 
This section provides details about the context of our study: i) the clinical scenario we focused on to address our research questions and ii) the data and apparatus. 

\subsection{Learning Scenario}
\label{sec:learning-scenario}
Our study was conducted within a third-year unit of the Bachelor of Nursing Program at Monash University. As part of their degree, students participate in high-fidelity clinical simulations designed to nurture teamwork and communication skills. Each class includes two consecutive simulations focused on prioritising care and managing clinical deterioration, structured as follows:
\begin{enumerate}
    \item[\textbf{1.}] \textbf{Pre-brief (10 minutes):} An in-class introduction to the upcoming scenario.
    \item[\textbf{2.}] \textbf{Clinical Scenario (20-30 minutes):} Students engage in a simulated healthcare scenario.
    \item[\textbf{3.}] \textbf{Post-scenario Debrief (30 minutes):} A class-wide debrief led by the teacher, focusing on team performance, communication, and prioritisation of patient care.
\end{enumerate}

Students are divided into teams of four, with two students acting as primary nurses and the other two as secondary nurses. The non-participating students observe the simulation via live stream. The clinical scenario is divided into three main phases: 
\begin{itemize}
    \item \textbf{Phase 1:} Two nursing students, playing primary ward nurses, enter the ward and receive \textbf{\textit{handover information}} for four patients. One of these patients begins to deteriorate, prompting the students to call for help.
    \item \textbf{Phase 2:} Nursing students, playing \textit{\textbf{secondary nurses, enter the scenario}} to assist in caring for the patients.
    \item \textbf{Phase 3:} Following a Medical Emergency Team (MET) call initiated by the students, the \textbf{\textit{doctor enters}} the ward to offer additional support.
\end{itemize}

\subsection{Multimodal Sensor-based Data and Apparatus}
\label{sec:multimodal-data}
We collected spatial and audio data streams using a \textbf{\textit{positioning indoor system}}\footnote{\url{https://www.pozyx.io}} and \textbf{\textit{wireless lapel microphones}}, respectively, for each team member participating in the clinical scenario. All the data is captured and synchronised with our multimodal data capture system \footnote{https://github.com/Teamwork-Analytics/teamwork-visualiser-dashboard}. Before the simulation begins, each student is given a belly bag containing a positioning sensor to track their \textit{x-y coordinates} and \textit{body orientation} within the simulation ward room (see Fig. \ref{fig:teaser} -- top) and is asked to wear a wireless lapel microphone to capture their \textit{audio}. Each student is also assigned a colour based on their role (i.e., blue and red for primary nurses -- PN1 and PN2 --, respectively, and green and yellow for secondary nurses -- SN1 and SN2 --, respectively), generating de-identifiable data streams. The simulation room was equipped with built-in video cameras, and an additional \textbf{180-degree camera} provided enhanced coverage. 
During the clinical scenario, an observer -- either a researcher or a teacher -- logs the \textit{main phases} to contextualise the multimodal data \cite{Huceta2022}. 
\section{Design of {\tool}}
\label{Design}
We followed a human-centred design (HCD) approach to iterate the design and development of \tool, focused on leveraging and repurposing AI techniques and solutions used in prior products or solutions. 
This approach, commonly used for designing innovations involving complex data and AI capabilities \cite{yangReexaminingWhetherWhy2020,Bly1999,Liao2023}, was well suited to the complexity and real-world scenario of our learning context.

In prior work, we conducted a feasibility study \textcolor{black}{(2021)} to explore the technological capabilities of capturing multimodal data during clinical scenarios, resulting in a dataset of team dynamics and initial analytics and AI models \cite{yan2023bjet,zhao2023LAK,zhao2023AIED}. 
Building on this, we developed a first high-fidelity prototype consisting of a set of initial visualisation prototypes that provided an overview of team dynamics, which were tested in an in-the-wild study where educators used them during debrief sessions \textcolor{black}{(2022)} \cite{Echeverria2024LAK,martinezmaldonado2023inthewild}. Feedback from these studies highlighted areas for improvement, leading to the second iteration presented in this paper: the design and implementation of {\tool}, a more flexible and dynamic system for visualising team dynamics during debriefings.

\subsection{Design Workshops to Elicit Educators' Needs and Requirements}
Five educators (1 simulation coordinator, 2 unit coordinators, and two lecturers; T1--T5) participated in the design of {\tool}. These educators (except T5) used the initial prototype in an earlier study. The design and validation of \tool~ consisted of three stages \cite{Martinez-Maldonado2015}: \textbf{(1) exploration of educators' needs and requirements, }\textbf{(2) low-fidelity prototyping}, and \textbf{(3) in-the-wild classroom study} (to be presented in Section \ref{sec:classroom-study}). 
\textcolor{black}{We conducted design workshops and testing sessions using low-, mid- and high-fidelity prototypes to iteratively shape the design of \tool, enabling educators to contribute to its functionality and interface.}
We conducted seven workshops with educators covering these three stages over five months \textcolor{black}{in 2023}. We decided to partner with educators to collectively reflect and discuss their current practices and engage in discussions leading to designing {\tool}'s features \cite{rosnerOutTimeOut2016,yangReexaminingWhetherWhy2020}. 
We used several artefacts to engage in the design of {\tool}, such as collaborative boards, wireframes, and low-fidelity prototypes\footnote{The following URL contains all the materials used during our design sessions \url{https://anonymous.4open.science/r/CHI2025-materials-3718/}}. Table \ref{tab:workshops} summarises the participants, goals and artefacts used during workshop sessions.

\begin{table*}[ht]
\small
\centering
\caption{Design workshops with the team consisting of researchers (R), developer (Dev) and educators (T).}
\label{tab:workshops}
\begin{tblr}{
  width = \linewidth,
  colspec = {Q[79]Q[160]Q[69]Q[167]Q[452]},
  cell{2}{1} = {r},
  cell{2}{3} = {r},
  cell{3}{1} = {r},
  cell{3}{3} = {r},
  cell{4}{1} = {r},
  cell{4}{3} = {r},
  cell{5}{1} = {r},
  cell{5}{3} = {r},
  cell{6}{1} = {r},
  cell{6}{3} = {r},
  cell{7}{1} = {r},
  cell{7}{3} = {r},
  cell{8}{1} = {r},
  cell{8}{3} = {r},
  cell{9}{1} = {r},
  cell{9}{3} = {r},
  cell{10}{1} = {r},
  cell{10}{3} = {r},
  cell{11}{1} = {r},
  cell{11}{3} = {r},
  hline{1,12} = {-}{0.08em},
  hline{2} = {-}{},
}
\textbf{Workshops} & \textbf{Participants} & \textbf{Duration} & \textbf{artefacts} & \textbf{Goal}\\
1 & R1-R4 & 60 mins & Collaborative board & {List all technology capabilities.\\List prior analytics and visualisations that could be potentially used, \\List of other tools and materials that educators use}\\
2 & T1, T2, R1, R2, Dev & 75 mins & Collaborative board & {Identification of educators' potential use of the features, analytics and visualisations. \\Co-creation and refinement of features from previous analytics and visualisations.}\\
3 & T3, T4, R1, R2, Dev & 63 mins & Collaborative board & {Identification of educators' potential use of the features, analytics and visualisations. \\Co-creation and refinement of features from previous analytics and visualisations.}\\
4 & T1 - T4, R1 - R4 & 66 mins & Low-fidelity prototype & Validation of a wired prototype of a control-view that shows visualisations and information\\
5 & T1, T2, T4, R1, R3, R4 & 64 mins & Low-fidelity prototype & Validation of wired prototype of an annotation tool\\
6 & T1, T2, T4, R1 - R4 & 60 mins & Low-fidelity prototype & {Definition of a list of observable events using video replay\\Co-design of a debrief-view to access and share visualisations with students.}\\
7 & T1-T5, R2 - R4, Dev & 120 mins & {mid-fidelity prototype\\low-fidelity prototype} & {Validation of annotation tool and refinement of observable events. \\Interaction design of the debrief-view to access and share visualisations with students.}\\
8 & T1, T2, T3, T5, R2 - R4, Dev & 120 mins & high-fidelity prototype & User testing of preliminary version using video replay from past clinical simulations\\
     \hline
\end{tblr}
\end{table*}

During most design workshops, all five educators were present, some being absent due to scheduling conflicts. All workshop sessions were screen and audio-recorded (via Zoom if the session was remote or via microphone if the session was in-person) and transcribed for analysis. After each workshop, two researchers convened to summarise the key points discussed, \textcolor{black}{and report the outcomes describing desired features, improvements, and suggestions. These outcomes were implemented and iteratively validated in subsequent workshops.} 

\paragraph{\textbf{(1) Exploration of educators' needs and requirements.}} 
In design workshops 1–3, we aimed to 1) gather educators' needs for supporting debriefing practices, 2) explore how they might use visual analytics to address interpretation challenges and enhance the design, and 3) refine or validate the system's features. Using a data-driven approach \cite{Gorkovenko2020futuredatadriven,Bly1999}, we generated visualisation ideas that were realistic and more relevant to the learning context, using a dataset from prior studies \cite{yan2023bjet,zhao2023LAK,martinezmaldonado2023inthewild}.
\textcolor{black}{Furthermore, considering the current proliferation of automated analytical processes driven by Generative AI (GenAI) tools and applications \cite{openai2024gpt4technicalreport}, we also validated ideas such as automatic identification of key moments using GenAI, similar to new features in commercial video annotation tools (see \href{https://anonymous.4open.science/r/CHI2025-materials-3718/}{\underline{supplementary material} - Section A 1.2}).} A collaborative board displayed visualisation examples such as priority charts, speech sociograms, and speech-location maps to spark conversations. Following a design critique technique \cite{Alabood2023Jan}, educators explored these visualisations and gave feedback on their potential use during debriefs. They also contributed to an empty wireframe, suggesting which visualisations they would prefer during debriefing. 
Adopting a thematic analysis, \cite{Braun_Clarke_2012}, two researchers analysed workshop transcripts, identifying 216 utterances and categorised these into 1) design considerations for using visual analytics in reflection and 2) design requirements for improving visualisations. These results are presented in Sections \ref{sec:design-considerations} and \ref{sec:data-and-visuals}, respectively.

\paragraph{\textbf{(2) Low-fidelity prototyping.}}
In workshops 4–7, we aimed to validate low-fidelity prototypes of \tool~ based on insights from previous sessions with educators. The goals were to 1) propose features for filtering, selecting, and combining visualisations, 2) design tagging and annotation features, and 3) define observable actions for educators.
In workshops 4 and 5, we used a collaborative tool \footnote{Figma -- \url{https://www.figma.com/}} to share wireframes, encouraging collaborative feedback on system features. In workshop 6, educators watched a video from a previous team simulation, annotated key events, and collectively agreed on observable actions, which were compiled into a spreadsheet. In workshop 7, they refined this list by tagging and annotating actions while watching another video.

Next, we present the design considerations (Section \ref{sec:design-considerations}) and visual analytics designs (Section \ref{sec:data-and-visuals}) resulting from the exploration of educators' needs. We then outline the system's interaction features and requirements derived from the low-fidelity prototyping workshops.

\subsection{Design Considerations}
\label{sec:design-considerations}

Based on the thematic analysis of educators' needs and requirements, we identified four design considerations (D1–D4) for integrating visual analytics into current debriefing practices.

\begin{enumerate}

\item[\textbf{D1:}] \textbf{Visual analytics should \textcolor{black}{facilitate constructive discussions on sensitive} topics (i.e., the elephant in the room):} Educators highlighted the challenge of addressing difficult topics, such as \textit{"things that didn't go well,"} without making students feel ashamed or singled out (T4). These topics often extend beyond clinical performance to include skills such as teamwork, communication, and prioritisation (T1). \textcolor{black}{While it is essential to discuss these aspects of student performance constructively, the visualisations should avoid presenting information that emphasises failures (T2).} The goal is to support comprehensive debriefs that balance sensitive feedback with constructive insights, addressing both clinical and interpersonal competencies while acknowledging their inherent complexities.

\item[\textbf{D2:}]\textbf{Visual analytics should provide more nuanced discussions:} Educators envision the use of analytics to foster constructive, more nuanced discussions, \textcolor{black}{exploring multiple aspects of team dynamics, such as communication patterns, task prioritisation, and role dynamics. }They highlighted the importance of analytics in providing an objective account of \textit{"the facts, the actions that happened"} (T4), thereby steering the conversation towards being \textit{"more informative"} (T4). Educators saw the potential for analytics to clarify specific areas requiring attention, such as instances of the lack of prioritisation, for example, where \textit{"nobody went over"} to a deteriorating patient (T2). This objective lens provided by analytics opens up space for discussions on challenges faced (T1) and essential skills needed, such as \textit{"teamwork, communication, prioritisation"} (T2). 

\item[\textbf{D3:}] \textbf{Educators should be able to customise the use of data and visualisations to adapt their discussions:} educators also acknowledged that each debriefing session is \textit{"structured differently"} (T1) depending on team's performance. Educators agreed that they would select relevant data points to focus on during the debrief (e.g., \textit{"a choice of four things that the data can give us"} -- T1) as not all visualisations would be useful in every scenario (T4). They could combine one or several visualisations to provide a comprehensive view of the team's dynamics because "each analytics provides a different point for discussion" (T4). The consensus is to initially integrate \textit{"one or maximum two"} pieces of data (T1), honing in on \textit{"the most important things"} (T3). This allows educators to customise the choice of visualisations to fit the team's needs. 

\item[\textbf{D4:}]\textbf{Educators should be able to tag, take notes and highlight key moments they observe during the learning activity to guide the discussion:} \textcolor{black}{Although the idea of automatically identifying key moments from the video and transcript was appealing, educators raised concerns about the potential inaccuracy of AI-generated outputs (T1, T3) and the time required to validate them, which make this approach unfeasible for their current practices (T1).} Thus, educators preferred to make live observations using a tagging and annotation tool, similar to \textcolor{black}{video annotation tools}, mirroring their existing practice of note-taking on paper (e.g., \textit{I always have a piece of paper, because I can write what I'm observing} -- T4). Educators \textcolor{black}{emphasised the importance of keeping full control by manually} annotating key moments of the simulation and bringing those into the discussion and reviewing the data and visualisations (e.g., \textit{tagging different points and then being able to recall them during the debrief} -- T1; \textit{link the tags with the video and the other data }-- T3). In this way, educators envision an efficient use of video snippets and visualisations to highlight key communication points and scenarios and quickly recap the team's performance over a short period, enhancing the understanding of team dynamics (T4). 

\end{enumerate}

\subsection{Data and Visualisations}
\label{sec:data-and-visuals}
This section presents the data and visualisations co-designed with educators to support debriefing practices. Addressing design consideration \textbf{--D2--}, the data and visualisations are aimed to allow educators to facilitate a nuanced discussion with three main objectives (derived from the learning outcomes of the simulation): 1) prioritisation strategies (i.e., task allocation), 2) teamwork dynamics (i.e., coordination) and 3) communication strategies (i.e., closed-loop communication, shared mindset). We report how educators interpreted these visualisations and the evolution of initial designs with direct inputs from educators.

\begin{figure*}[ht]
\includegraphics[width=\textwidth]{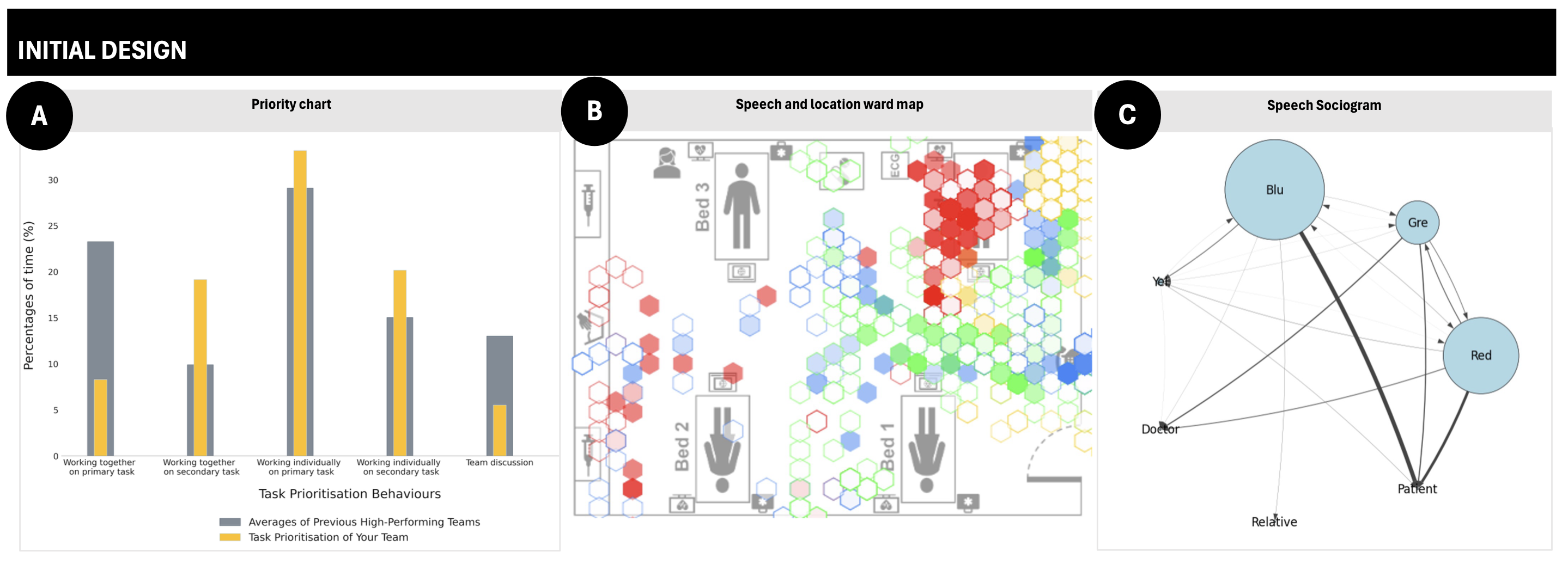}
\caption{\textcolor{black}{Initial designs considered during the workshop design sessions. A) a priority chart depicting the team's prioritisation behaviours (orange bar) and the values from previous high-performing teams. B) Speech and location ward map illustrating the amount of speaking time in a particular location. C) Speech sociogram showing the amount of speaking time per student and their interactions with other students/roles. }}
\label{fig:evolution}
\Description{Three visualisations: a bar chart, a density map with hexagons, and a sociogram. }
\end{figure*}

\subsubsection{Priority chart:} 
A \textit{bar chart} represented six team prioritisation behaviours based on team members' locations (x-y coordinates). These behaviours capture whether students prioritised care for bed 4 (primary task) or other beds (secondary task). From this, we calculated collaborative and individual work on both tasks, as well as transitions between beds. These behaviours have been evaluated in a prior study \cite{yan2023bjet}. The initial \textit{bar chart} design showed a grey bar indicating the cumulative percentage of time spent on each behaviour and a yellow bar representing the average time spent by high-performing teams (see Fig. \ref{fig:evolution}- Initial design \itemA ). 
\textit{Bar chart} facilitated educators to explain that teams are expected to spend more time on the primary task, especially when the patient is deteriorating, allowing them to discuss prioritisation behaviours. For underperforming teams, the chart provided evidence to facilitate constructive dialogue, allowing comparison of their performance with high-performing teams.

However, based on their experience with the initial design \cite{Echeverria2024LAK}, educators requested to remove the percentage of the high-performing teams' average (Fig. \ref{fig:evolution} -- Initial design \itemA - grey bar), as students might feel uncomfortable when comparing their performance with that of high-performing teams. They emphasised that simulations are meant to be a safe learning environment. Students already feel pressure to perform well in high-fidelity clinical settings and want to avoid negative emotions during the debrief.
Therefore, with this feedback, we show only the current team's behaviours in the bar (Fig. \ref{fig:evolution} - \itemA).
An additional improvement was to use more representative labels such as \textit{"working together for the main patient"} and \textit{"working together on non-critical tasks [attending other patients]"}.

\begin{figure*}[ht!]
\includegraphics[width=\textwidth]{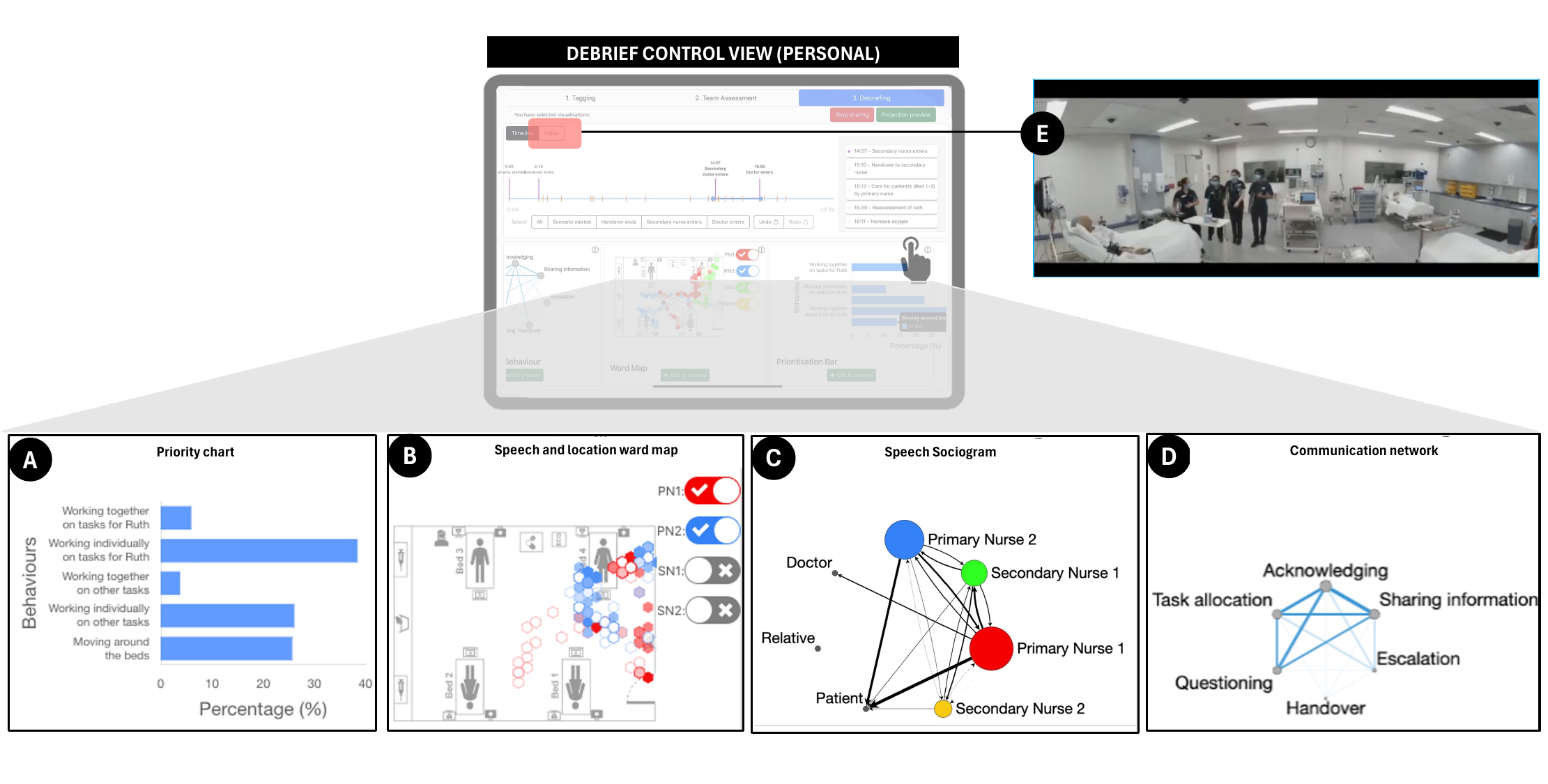}
\caption{Catalogue of visualisations and sources in the debrief control view: A) Priority chart, B) Speech and location ward map, C) Speech sociogram, D) Communication network, and E) Video snippet.}
\label{fig:visualisations}
\Description{Figure A: Priority Chart. This bar chart compares task prioritisation behaviours between high-performing teams (grey bars) and the user's team (yellow bars) in a clinical simulation setting.
X-axis: Represents four task behaviours—working together on the primary task, working together on the secondary task, working individually on the primary task, and engaging in team discussions.
Y-axis: Indicates the percentage of time spent on each task behaviour.
Key Insights: High-performing teams allocated more time to team discussions and working together on the secondary task, whereas the user's team spent a larger proportion of time working individually on the primary task.
Figure B: Speech and Location Ward Map. This figure shows a spatial map of a clinical ward, overlaid with a heatmap of speech and movement activity. The heatmap is colour-coded and also allows to depiction of the quantity of speech with fades. A filled hexagon means a great quantity of speech, while a transparent hexagon means a lack of speech.
Layout: The ward contains labelled beds (e.g., Bed 1, Bed 2, Bed 3) and various medical equipment.
Purpose: Visualizes the distribution of verbal interaction and movement within the ward to analyse team dynamics during the simulation.
Figure C: Speech Sociogram. This sociogram visualizes speech interactions among team members during the clinical simulation.
Nodes: Represent individuals in the simulation (e.g., "Blu," "Red," "Gre," "Patient," "Relative"). Node size corresponds to the volume of speech contributions.
Edges: Show the direction and frequency of speech interactions between individuals. Thicker edges indicate more frequent interactions.
In this example, "Blu" is the most active participant, as indicated by the large node and thick connections to others, particularly "Red" and "Patient."
}
\end{figure*}

\subsubsection{Speech and location ward map:} 
To visualise spatial and audio data, we use a Hexbin map\footnote{\url{https://d3-graph-gallery.com/hexbinmap.html}} to combine both data sources in one visualisation. Inspired by sports analytics and spatial visual analytics \cite{Perin2018sportdata,goldsberry2012courtvision}, this visualisation resembles a density map, but instead of representing each data point's position individually, we used a binary indicator to show the presence or absence of audio (see Fig. \ref{fig:evolution} -- Initial design \itemB ). 
The \textit{x-y} coordinates were scaled and rendered according to the size of a layout image that was used to represent the physical learning space. Each data point is mapped into a hexagon, representing a small area in the ward map. Each student is assigned a colour (primary nurses: blue and red; secondary nurses: green and yellow). We use colour intensity to render the presence/absence of communication, computed using a Voice Activity Detector (VAD) algorithm. For instance, if the hexagon has a \textit{filled colour}, it means that the student was actively speaking most of the time while standing in that particular position, whereas if the hexagon has a \textit{white fill}, the student located at that particular position was not speaking. 

From their prior use, educators found the \textit{ward map} helpful in guiding their discussions on communication, teamwork and task allocation strategies at an individual level and as a team. They envisaged using the \textit{ward map} to explore key moments in the simulation, such as a nurse's movement about their prioritisation strategies. 
Educators would expect primary nurses to mainly handle the main patient (bed 4) or escalate the situation by calling the Medical Emergency Team (MET), focusing less on covering beds 1, 2, and 3. In contrast, secondary nurses should mostly cover secondary tasks (beds 1, 2, and 3) and help primary nurses with the primary task (bed 4).
Educators asked about the possibility of filtering the information by particular periods so they could see differences in students' teamwork strategies. This was mainly discussed because the initial prototype illustrated cumulative data, which made it hard for educators to get particular insights or moments for discussion.

\subsubsection{Speech sociogram:} 
Two different visual analytics were presented to educators to illustrate speech interaction: a \textit{sociogram} \cite{kim2008meetingmediator,gasevic2019sens}, which has been adopted in both face-to-face \citep{echeverria2019collaboration} and online \citep{zhou2021investigating, Praharaj2022collaborativeconvergence} learning scenarios and a timeline of speech presence (e.g.,\cite{Samrose2021}). One particular visual analytics that educators found helpful was the sociogram \cite{echeverria2019collaboration, zhou2021investigating}. Educators ruled out having a timeline with speech presence/absence, as it is difficult to quickly grasp interactions if two students talk simultaneously or to each other.  
Fig.\ref{fig:evolution} -- Initial design \itemC -- depicts the initial design of the sociogram. The nodes in the \textit{sociogram} depict team members, doctors, and relative speaking time. The arrows (directed lines) depict the verbal interaction and direction of communication between each individual, and the arrow's thickness represents the amount of time they interacted with each other. The speaking time is calculated using a Voice Activity Detector (VAD) algorithm\footnote{py-WebtcVAD: \url{https://github.com/wiseman/py-webrtcvad}} from individual speech data. The direction of communication is calculated by inferring an f-formation using orientation and x-y coordinates \cite{Kendon2010}. Examples of such f-formations include the participants standing face-to-face or shoulder-to-shoulder in close proximity \cite{zhao2022modelling}. 

The \textit{sociogram} proved particularly useful for discussing each student's communication contribution, as it visualises who communicated with whom and how much at specific times. Educators agreed it helps facilitate discussions on critical healthcare practices, such as patient care and teamwork, including recognising and interacting with other roles in the multidisciplinary team. Additionally, educators noted that the sociogram could guide discussions on team coordination, including who assumed leadership roles and how tasks were allocated.
Concerning the improvements of the \textit{sociogram}, educators wanted individuals to be identified by colours. Therefore, we assigned a colour to each node according to the student's role, as depicted in Fig.~\ref{fig:visualisations} -- \itemC~ (PN1 = red, PN2 = blue, SN1 = green, SN2 = yellow).

\subsubsection{Communication network:} 
\label{sec:design-ENA}
As guided by healthcare frameworks \citep{haig2006sbar, hargestam2013communication}, communication is highly contextual. Previous studies have found such contextual communication can be captured and analysed through epistemic network analysis \citep{zhao2024lak, swiecki2020assessing}, which is a relatively novel approach to analyse the co-occurrence between codes in \textcolor{black}{qualitative} data \citep{shaffer2016tutorial}. Thus, we adopted the \textit{epistemic network} to capture the communication behaviours for each team. From existing theories on teamwork and healthcare \citep{miller2009identifying,riley2008nature,alonso2012building,bigfive}, we used six communication behaviours \textcolor{black}{to code the dialogue content}: 1) acknowledging, 2) sharing information, 3) questioning, 4) task allocation, 5) handover, and 6) escalation. In the \textit{epistemic network}, nodes represent these six codes, lines show the co-occurrence of two codes (i.e., one code happening after another), and edges reflect the frequency of co-occurrences. 
To create this epistemic network, students' speech data were transcribed using OpenAI Whisper \citep{radford2023robust} and coded with a fine-tuned BERT-based model \citep{BERT}. Specifically, this BERT-based model was fine-tuned with the coded transcriptions that were previously collected in the same clinical scenario. These transcriptions were manually coded with the six communication behaviours, creating a dataset containing 3,647 sentences for fine-tuning. The performance of the model can be found in \cite{Zhao2024BJET}.

When educators interacted with the \textit{communication network}, they found it helpful in showing the content of the communication but in a compact manner, eliminating the need to access the full transcript. 
Educators also seemed interested in how these codes could help understand team interactions during the simulation more deeply. For example, educators mentioned that \textit{"task allocation"} and \textit{"handover"} reflect how tasks and patient information are distributed among team members, while \textit{"questioning"} and \textit{"responding"} demonstrate interactive dialogue that supports decision-making. Educators also discussed how \textit{"acknowledging"} could be used to demonstrate both effective leadership and teamwork, depending on the situation. For example, a higher frequency of \textit{"acknowledging"} might indicate strong teamwork and responsiveness.
Given that the current \textit{communication network} provides an information summary of the whole scenario, educators suggested allowing filtering by key moments and phases of the simulation to provide contextual insights. 

\begin{figure*}[ht!]
\includegraphics[width=\textwidth]{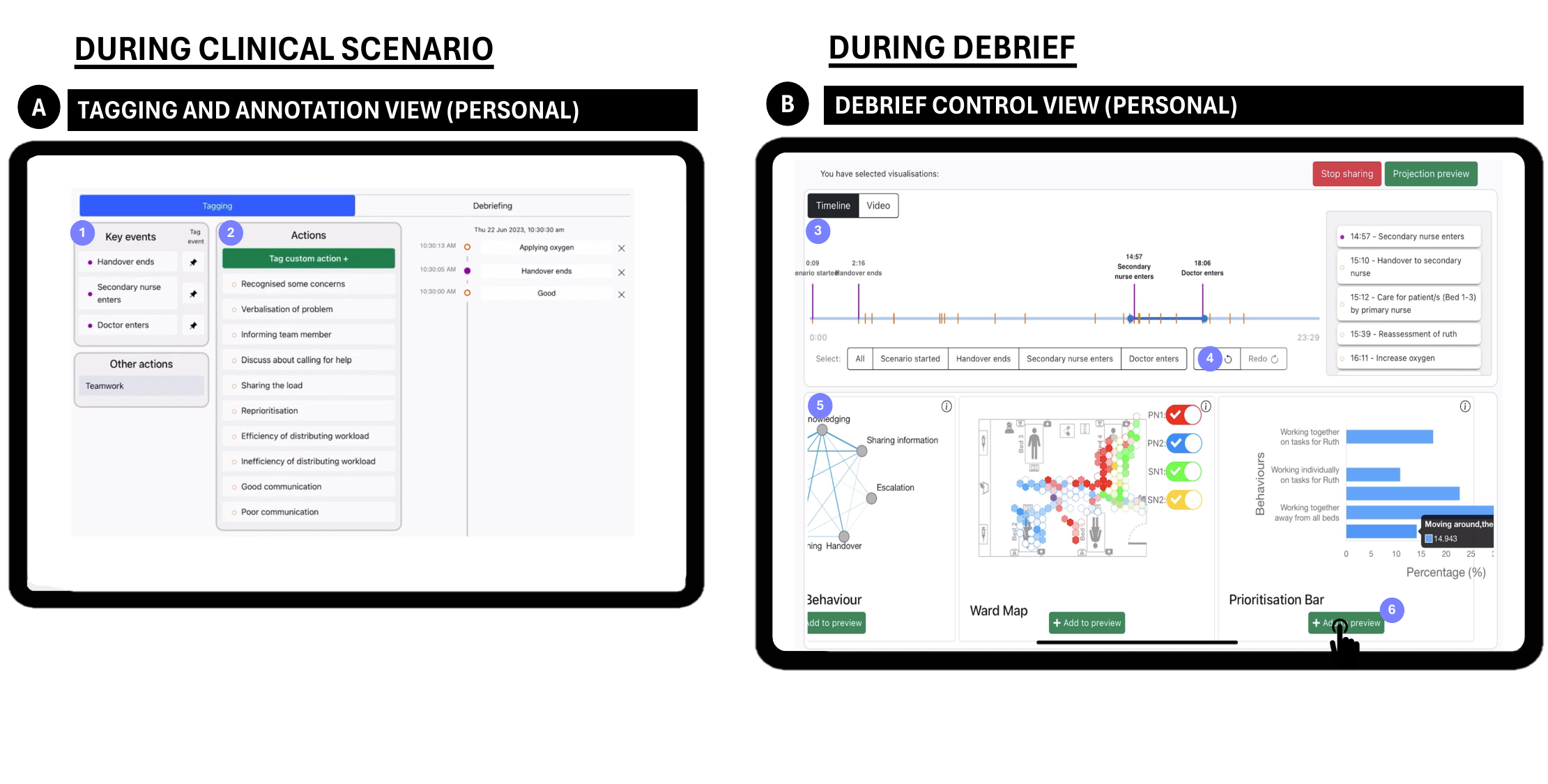}
\caption{(\textcolor{black}{A) Tagging and annotation view at the personal device during observation of clinical scenario. Educators can choose from (1) main phases or (2) actions. B) Debrief control view on the personal device. The main features are (3) the timeline to control the selection and navigation of data, (4) a quick selection of main phases, (5) a catalogue of visualisations, and (6) the option to share a visualisation to the shared screen. }}
\label{fig:three-screens}
\Description{Figure A: Tagging and Annotation View (Personal).
This figure shows the "Tagging and Annotation View," designed for individual use during a clinical scenario. The interface includes key components for tagging and annotating events in real-time. The view is structured to allow users to record observations during a clinical simulation scenario. Key Features: Multiple annotation tools are present, enabling the tagging of significant events. Sections for written notes and predefined tags are displayed, allowing for quick categorization. Purpose: This view supports the user in creating detailed, structured feedback during later discussions in debriefing sessions. Figure B: Debrief Control View (Personal)
This figure illustrates the "Debrief Control View," used individually during a debriefing session. Overview: It provides tools and visualizations for reviewing tagged and annotated data collected during the clinical scenario.
Key Features: A timeline or list of tagged events for chronological review.
Controls for revisiting specific points in the simulation for discussion.
Purpose: This view guides debriefing discussions by offering an organized presentation of previously recorded events and notes.}
\end{figure*}

\subsection{{\tool}'s Interface}
\label{sec:interface}
\textcolor{black}{Considering educators' inputs from the ideation and testing sessions with low-, mid- and high-fidelity prototypes,} we designed and implemented three views: A) a tagging and annotation view, B) a debrief control view, and C) a shared screen view (See Fig. \ref{fig:three-screens}). The \textit{tagging and annotation view} and the \textit{debrief control view} were designed to be used on a tablet or a personal device, as educators wanted to have portability and ease of access when using the system.

\begin{enumerate}
    \item [\textbf{A.}]\textbf{Tagging and annotation view:}
Addressing design consideration \hyperref[sec:design-considerations]{\textbf{D4}}, where educators emphasised the need to tag, annotate and highlight key moments manually, this personal view is used when observing the simulation and provides a list of actions allowing 
educators \textit{\textbf{to tag predefined actions}} and phases during live observations (See Fig. \ref{fig:three-screens} \itemA). Educators recognised two main categories for tagging. One category is related to the \textit{main phases} or key actions: \phone~ \textit{handover ends}, \phtwo~ \textit{~secondary nurse enters} and \phthree~ \textit{doctor enters}, which were derived from the learning design. The second category corresponds to \textit{actions} expected to occur in each phase. 
In addition, educators can \textit{\textbf{annotate their observations}}, which are linked to a particular action. 

 \item [\textbf{B.}]\textbf{Debrief control view:}
Once the clinical simulation ends, educators can access a debrief control view to manage the information they share with students. This "all-in-one" personal view allows educators to navigate and select phases (timeframes) and actions for debriefing, addressing design consideration \hyperref[sec:design-considerations]{\textbf{D3}} (see Fig. \ref{fig:three-screens} \itemB). A \textbf{\textit{timeline}} with four phases -- \textit{all, handover ends, secondary nurses enter, and doctor enters} -- anchors the video and visualisations; it also displays actions annotated by the educator. Selecting a phase or action dynamically updates the visualisations and indexes a 20-second video snippet.

Educators can choose visualisations from the \textbf{\textit{catalogue of visualisations}}  (as described in Section \ref{sec:data-and-visuals}), and they can \textit{add to projector}, meaning that the visualisation will be shared with all students in the room (displayed in a shared screen). Educators can also stop sharing the visualisation or opt to share and playback the video snippet.  

\item [\textbf{C.}]\textbf{Shared screen view:}
This shared view aims to simplify the information display by showing specific elements educators deem important (i.e., video or visualisations) to avoid overwhelming students with all the information at once, addressing design consideration \hyperref[sec:design-considerations]{\textbf{D1}} (see Fig. \ref{fig:teaser} \itemC). Educators can show in the shared screen one to three visualisations at the same time.
\end{enumerate}

To implement \tool\footnote{A demo of \tool~ can be found in  \url{https://teamwork-analytics.github.io/dashboard-ghpage/}}, we used a WebSocket\footnote{\url{https://socket.io/}} connection between the projector screen and the handheld device. This connection allowed the projector to listen for the selected element(s) from the handheld device and display the corresponding visualisation in real-time.

\textcolor{black}{
To summarise the design process, Table \ref{tab:design-decisions} maps the design considerations to the corresponding design decisions and their implementation as \tool's features. The first four rows outline the design considerations identified during the design workshops (\hyperref[sec:design-considerations]{D1 -- D4}), while the last three rows describe the specific improvements to data and visualisations requested by educators (\hyperref[sec:data-and-visuals]{I1 -- I3}).}

\begin{table*}[ht]
\caption{\textcolor{black}{Mapping design considerations, educators' design decisions, and their translation into \tool's features.}}
\label{tab:design-decisions}
\resizebox{\textwidth}{!}{%
\renewcommand{\arraystretch}{1} 
\begin{tabular}{@{}lll@{}}
\toprule
\textbf{Design considerations} &
  \textbf{Educator's design decisions} &
  \textbf{TeamVision features} \\ \midrule

\begin{tabular}[t]{@{}l@{}}
D1 -- Facilitate constructive discussions \\ on sensitive topics
\end{tabular} &
  \begin{tabular}[t]{@{}l@{}}
  Review and control which visualisations \\ 
  are shared with students during debriefing
  \end{tabular} &
  \begin{tabular}[t]{@{}l@{}}
  \textbf{Debrief control view:}\\     
  Share selected visualisations from personal debrief \\ 
  view to a shared view (projector)
  \end{tabular} \\ \hline

\vspace{0.25cm} 

D2 -- Provide more nuanced discussions &
  \begin{tabular}[t]{@{}l@{}}
  Have access to information that allows \\ 
  educators to explore multiple aspects of \\ 
  team dynamics and compare how these \\ 
  dynamics evolve over time 
  \end{tabular} &
  \begin{tabular}[t]{@{}l@{}}
  \textbf{Catalogue of visualisations:} \\      
  \textbf{Priority chart} summarises team's priority \\ 
  behaviours (group level).\\      
  \textbf{Speech and location ward map} shows speech \\ 
  and movement (individual level).\\      
  \textbf{Speech sociogram} shows non-verbal communication \\ 
  patterns (individual level).\\      
  \textbf{Communication network} summarises dialogue \\ 
  content (group level).
  \end{tabular} \\ \hline

\vspace{0.25cm} 

\begin{tabular}[t]{@{}l@{}}
D3 -- Support customisation of data and \\ 
visualisations
\end{tabular} &
  \begin{tabular}[t]{@{}l@{}}
  Filter data by simulation phases or moments \\ 
  and allow educators to choose specific \\ 
  visualisations
  \end{tabular} &
  \begin{tabular}[t]{@{}l@{}}
  \textbf{Debrief control view:}\\      
  Real-time data filtering through the selection of \\ 
  phases or periods in an interactive timeline.\\      
  Selection of visualisations (up to three) from a catalogue.
  \end{tabular} \\ \hline

\vspace{0.25cm} 

\begin{tabular}[t]{@{}l@{}}
D4 -- Support manual tagging, annotation, \\ 
and highlight of observations
\end{tabular} &
  \begin{tabular}[t]{@{}l@{}}
  Create a tagging feature inspired by video \\ 
  annotation tools, allowing educators to \\ 
  annotate predefined actions and phases during \\ 
  live observations and mark key actions.
  \end{tabular} &
  \begin{tabular}[t]{@{}l@{}}
  \textbf{Tagging and annotation view:}\\      
  Pre-defined actions.\\      
  Add notes on each action.\\      
  Mark key actions (favourites).
  \end{tabular} \\ \hline

\vspace{0.25cm} 

\begin{tabular}[t]{@{}l@{}}
I1 -- Effectively represent team communication \\ 
dynamics, making patterns and roles \\ 
easier to interpret.
\end{tabular} &
  \begin{tabular}[t]{@{}l@{}}
  Refine the sociogram by adding role-based \\ 
  colour coding to make communication patterns \\ 
  more visually distinct and easier to interpret.
  \end{tabular} &
  \begin{tabular}[t]{@{}l@{}}
  A \textbf{sociogram} with role-based colour coding \\ 
  (primary nurses: red and blue; secondary nurses: \\ 
  green and yellow).
  \end{tabular} \\ \hline

\vspace{0.25cm} 

\begin{tabular}[t]{@{}l@{}}
I2 -- Facilitate access to key communication \\ 
strategies, avoiding the need to review \\ 
transcripts.
\end{tabular} &
  \begin{tabular}[t]{@{}l@{}}
  Add information that summarises key \\ 
  communication themes and strategies.
  \end{tabular} &
  \begin{tabular}[t]{@{}l@{}}
  A \textbf{communication network} that provides a summarised \\ 
  view of dialogue content (nodes are codes, edges \\ 
  are co-occurrences of codes).
  \end{tabular} \\ \hline

\vspace{0.25cm} 

\begin{tabular}[t]{@{}l@{}}
I3 -- Avoid features that could demotivate \\ 
students.
\end{tabular} &
  \begin{tabular}[t]{@{}l@{}}
  Remove priority chart comparisons to \\ 
  high-performing teams.
  \end{tabular} &
  A \textbf{priority chart} summarising five priority behaviours. \\ \bottomrule
\end{tabular}
}
\end{table*}

\section{In-the-wild Classroom Study}
\label{sec:classroom-study}
We conducted an in-the-wild classroom study. In-the-wild studies have been adopted at a larger scale and uncover the usage and impact of new technologies in non-controlled conditions \cite{Crabtree2013innovationinthewild,Rogers2007hassle,Chamberlain2012inthewild}. Educators used {\tool} in post-scenario debriefs to discuss team performance with students. The study, approved by the Monash University Human Research Ethics Committee, took place over one month in the second semester of 2023. One week before commencing the classroom study, all educators participated in a training session to become familiar with the tool.

\subsection{Data Collection}
We collected spatial and audio data using our multimodal data collection system (Section \ref{sec:multimodal-data}). Capturing these data is considered non-intrusive and consistent with standard practices in nursing simulations \cite{seropian2010design}. 
We followed an opt-out approach, meaning that {\tool} was used as \textit{"business-as-usual"} and that students could opt out to remove their data from the study. 
The visualisations shown in {\tool} (see Fig. \ref{fig:visualisations}) were automatically generated and used by educators in the debriefs without the researchers' involvement. The study included 7 educators and 221 students from 56 teams. Two researchers were on-site to address technical challenges.

\subsubsection{Audio recordings from debrief sessions.}
Educator debriefings while using \tool~ were audio-recorded. Educators consented to be audio-recorded and wore a lapel microphone to capture their speech. Due to missing \textcolor{black}{ data or }audio during the data collection or debrief sessions, \textcolor{black}{30} teams were excluded. Thus, we transcribed and analysed 26 sessions (SID 1--26). 

\subsubsection{Interaction data from~\tool~during debrief sessions.}
In addition, we captured interaction data (i.e., clicks and timestamps) from \tool~ to understand how educators interacted with the system. We also video-recorded the shared screen view to access the visualisations educators used during their debriefing sessions. 

\subsubsection{Post-hoc study with students.}
Students who participated in the simulation were invited to attend a one-hour follow-up interview to investigate their perceptions in terms of usefulness, trust and accuracy of the visualisations that the educators used during the debrief. Fifteen students (referred to as S1--S15; 4 males and 11 females, avg. age: 21.85, std. dev: 2.24) participated and were rewarded with a \$40 gift card. During the interview, students explored their own visualisations to provide more context to their responses. All interviews were held through a video-conferencing system (i.e., Zoom), video-recorded and transcribed for analysis. 

\subsubsection{Post-hoc study with educators.}
One month after the data collection, the five educators (T1--T5) were asked to participate in a one-hour follow-up interview to reflect on the use of \tool. In this interview, educators were invited to share their thoughts about the support of \tool~ for debriefing, the challenges they faced (if any), and their perceptions of the usefulness, trust and accuracy of the visualisations. All interviews were held through a video-conferencing system (i.e., Zoom), video-recorded and transcribed for analysis. 

\subsection{Analysis}

For \textbf{RQ1}, we aimed to assess how \tool~ supported tailored debriefing by analysing how educators adapted their practices. Since each scenario phase represented different tasks (See Section \ref{sec:learning-scenario}) we structured our analysis around these phases. Using interaction data and video recordings from the shared screen, we annotated the phases and visualisations used in each session. We then examined the frequency and combination of visualisations employed across phases, identifying patterns that revealed which visualisations were preferred during specific phases.
We also explored the topics covered by these visualisations. We split debriefing transcripts into 655 utterances and conducted a thematic analysis using an inductive approach \cite{Braun_Clarke_2012}. Two researchers independently reviewed and annotated the same five sessions, collaboratively defining sub-themes and high-level themes. The remaining sessions were divided for further analysis \cite{mcdonald2019reliability}. To quantify the thematic data, we calculated a coverage index by dividing the frequency of each theme by the number of sessions where it appeared. This enabled deeper insights into patterns of visualisation use and the associated discussions.

For \textbf{RQ2}, we conducted a thematic analysis of post-hoc interview transcripts, using a deductive approach to identify perceived benefits and challenges \cite{Braun_Clarke_2012}. Instances from the interviews were grouped into key topics through affinity diagramming, highlighting the benefits and challenges educators encountered when using \tool.

For \textbf{RQ3}, we applied a similar thematic analysis to post-hoc interviews with educators and students, focusing on their perceptions of usefulness, trust, and accuracy. Instances were categorised based on these two themes, and results were summarised per visualisation. 

Two researchers collaboratively analysed the topics from \textbf{RQ2} and \textbf{RQ3}, discussing findings regularly to ensure consistency and alignment \cite{mcdonald2019reliability}.

\section{Results}
\subsection{{\tool} Strategies and Use to Support Debriefing}
\label{sec:results-RQ1}

\subsubsection{Identified strategies}
\label{sec:results-RQ1-strategies}
\begin{figure*}[ht!]
\includegraphics[width=\textwidth]{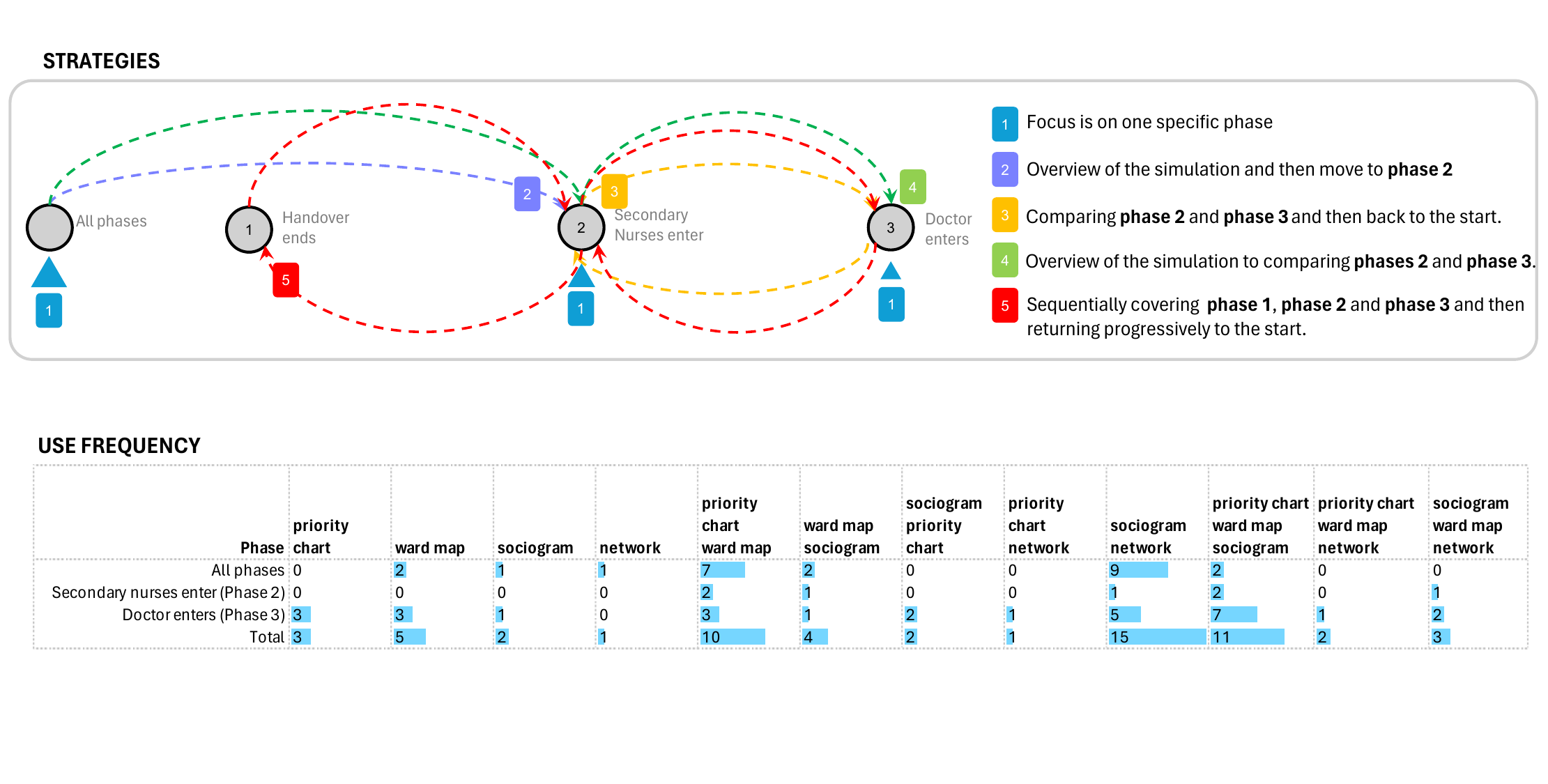}
\caption{\textbf{Top: }Strategies followed by educators while using the filtering option during their debrief sessions. Note how educators mostly used only the information of one particular phase to generate their corresponding visualisations and support their discussions. \textbf{Bottom:} Use frequency of visualisations per phase. The \EN~ and \SNA~ were frequently used together when visiting \textit{all} phases.}
\label{fig:strategies}
\Description{Top figure: Flowchart illustrating four strategies with pathways and use frequency chart for priority, ward, sociogram, and network maps across different phases and entries.
Bottom: mini bar charts depicting the use frequency of different combinations of visualisations during the use of the tool and according to the phases of the simulation.}
\end{figure*}

 We identified five distinct strategies based on analysing how educators utilised the filtering options (phase selection) from \tool. 
 Fig. \ref{fig:strategies} illustrates the five different strategies educators followed when using \tool. 
\textbf{Strategy 1} was the most common approach, with educators focusing on a specific phase of the clinical simulation (see Fig. \ref{fig:strategies} - top). In 10 sessions, they solely used the \allphases phases filter to discuss the data summarising key aspects of the simulation. Additionally, five sessions focused on phase \phtwo \textit{ -- secondary nurses enter}, while two sessions focused on phase \phthree -- \textit{doctor enters}. This suggests educators prioritised certain phases based on their instructional goals or areas they wanted to emphasise during the debrief but could not compare and contrast information from other phases.

Educators also employed dynamic strategies, comparing multiple phases during the whole debriefing session (Fig. \ref{fig:strategies} - top). 
In \textbf{strategy 2}, they visited \allphases phases and then focused on specific points from phase \phtwo~  (e.g., SIDs 8 \& 9).
In \textbf{strategy 3}, they used visualisations to progressively compare phases \phtwo~ $\rightarrow$ \phthree~ $\rightarrow$ \phtwo~ (e.g., SIDs 7 \& 10).
 \textbf{Strategy 4} compared \allphases $\rightarrow$ \phtwo $\rightarrow$ \phthree, starting with an overview in \allphases phases, then analysing team dynamics in \phtwo, and concluding with a key moment from the video in \phthree (only SID 22).
Finally, \textbf{strategy 5} represents a more complex approach, observed in two debrief sessions (SID 23 \& 26), where educators switched between phases for comparison and then revisited earlier phases to recap and conclude the discussion (Fig. \ref{fig:strategies} - top).

Overall, these nuanced strategies illustrate diverse approaches to using the phase-filtering options, with a strong inclination toward a comprehensive review (\allphases phases) or a focused analysis on specific phases. Next, we revisit how educators used the visualisations in each phase. 

\subsubsection{Frequency of visualisations' usage}
\label{sec:results-RQ1-frequency}
To examine how educators used visualisations during debriefs, we analysed the frequency of visualisation combinations across simulation phases. The \textcolor{black}{bar charts}  (see Fig.  \ref{fig:strategies} - bottom) reveal that the combination of \EN~ and \SNA~ was the most common, especially in \allphases~ (9 occurrences) and phase \phtwo~ ~(4 occurrences), suggesting educators relied on these communication-based visualisations to reflect on team dynamics. The combination of \SNA, \barchart, and \wardmap~ was also frequently used in phase \phtwo(7 occurrences) to discuss task allocation and teamwork. The \barchart~ and \wardmap~ combination was often used in \allphases~ phases (7 occurrences).
Less frequent combinations, such as the \SNA, \barchart~ or \wardmap~ alone, as depicted in Fig.  \ref{fig:strategies} - bottom (first three columns), indicate educators preferred integrating multiple data sources to support a comprehensive reflection.

Next, we present the results of our qualitative analysis organised by topics that offer further explanations of the most commonly used combinations of visualisations.

\subsubsection{Themes discussed during debriefs}
\label{sec:results-RQ1-themes}
\begin{figure*}[ht!]
\includegraphics[width=\textwidth]{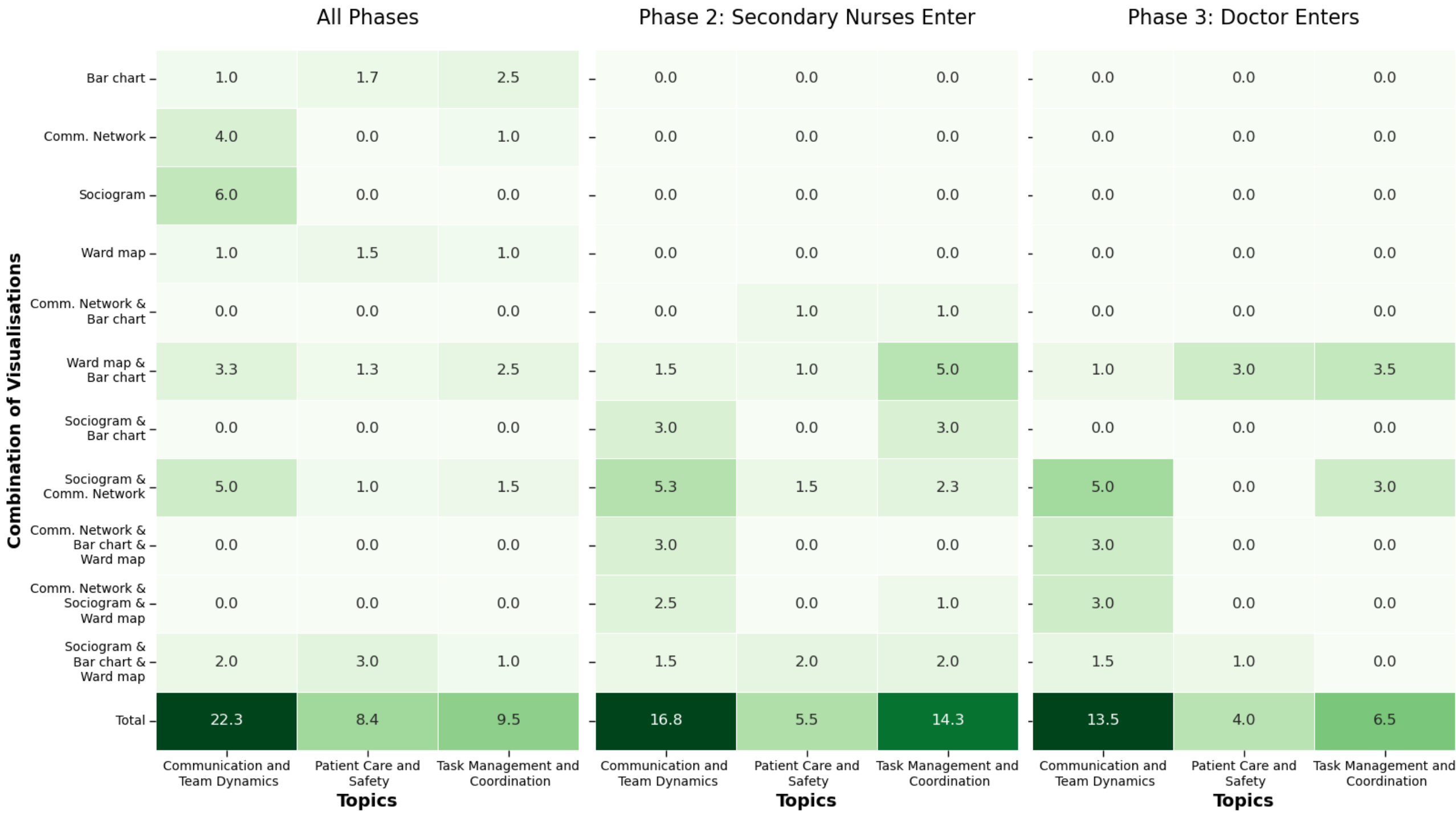}
\caption{Topic coverage index per combination of visualisations. }
\label{fig:coverage}
\Description{three heatmap charts displaying various phases related to the clinical scenario: the first is for all phases, the second is for phase 2 when the secondary nurses enter into the simulation, and the third is for phase 3 when the doctor enters into the simulation. Each heatmap depicts the most relevant topics covered during the discussion session, led by teachers.
X-axis: three main topics, which are communication and team dynamics; patient care and safety; and task management and coordination.
Y-axis: the combination of visualisations used during the discussion to cover the different topics, for example: bar chart and ward map, bar chart and sociogram, and so on. 
Each point on the heatmap corresponds to the topic coverage, which means the ratio of topic covered for a particular combination of visualisations.
}
\end{figure*}

Using the topic coverage, as observed in Fig. \ref{fig:coverage}, we discuss the most prominent topics covered per each phase. 

\paragraph{\textbf{Communication and team dynamics:}}
The combination of \EN~ and \SNA~ provided a rich and nuanced picture of team dynamics by highlighting both the content and the structure of communication. Educators mostly focused on discussing communication strategies such as \textit{\textbf{"closed-loop communication"}} and team dynamics in the form of \textit{\textbf{"multidisciplinary communication"}}. 

For example, educators used \EN~ to highlight instances of closed-loop communication, where tasks were allocated, acknowledged, and confirmed as complete. This type of communication is critical for ensuring clarity in emergency responses. As one educator stated: \textit{"When we acknowledge something needs to be done, we have clarifying questions, we allocate a task, a task is completed, and that task is acknowledged as complete. Closed loop."} (SID 6, \allphases). Then, using the \SNA~, educators discussed multidisciplinary communication by highlighting effective communication (nurse-nurse, nurse-doctor, nurse-patient): \textit{"We've got lots of communication in other directions...So, really good communication around that."} (SID 6, \allphases).
Subtle differences were observed in discussions when using \SNA~ in different phases. In phase \phtwo, educators focused on specific team interactions, such as nurse-nurse or nurse-patient/relative communication: \textit{"And then if we have a little look there again, you're both communicating nice and clearly and the patient as well"} (SID 5). In phase \phthree, educators highlighted nurse-doctor interactions: \textit{"We've got communication between all our nurses...But particularly with the doctor as well"} (SID 1).

\paragraph{\textbf{Patient care:}}
The combination of the \wardmap, \barchart, and \SNA~ allowed educators to discuss patient care and emphasise the need for balanced \textbf{"\textit{task allocation}"}, \textbf{"\textit{clear communication}"} and effective \textit{\textbf{"spatial awareness"}} when multiple patients required attention. 

While the \barchart~ provided an overview of task focus (other patients or the main bed), the \wardmap~ showed how the team was physically positioned to provide care. The \SNA~ clarified the communication and coordination efforts behind effective patient care. Educators highlighted how nurses managed the emergency at the main bed using the \barchart~ and then used the \wardmap~ to pinpoint tasks and nurse positions. One educator explained: \textit{"We realised that Ruth was a high priority, and we've dealt with that...like [yellow nurse], you came in, and you were sort of at that side of the bed, changing oxygen, helping with the bolus"} (SID 25, \allphases).

The combination of the \wardmap~ and \SNA~ supported discussions on communication structure, showing how well the team maintained coordination while caring for patients at both the primary and other beds: \textit{"Look at you both talking to patients. So, those patients would have felt reassured and cared for"} (SID 9, \allphases). Both visualisations also revealed that nurses were located around the deteriorating patient (bed 4) and strategically arranged to support communication and assist other patients (SID 8, phase \phtwo).

The combination of the \wardmap~ and \barchart~ helped educators confirm the coordination and distribution of nurses during the emergency, emphasising the importance of caring for both the main bed and other patients: \textit{"[student name] when you were moving around caring for the other patients and like we already identified, a really important thing to do, and particularly because you had a lot of other stuff going on as well."} (SID 1, phase \phthree).

\paragraph{\textbf{Task management and coordination:}}
The combined use of the \barchart~ and \wardmap~ was relevant for facilitating discussions on \textit{\textbf{"task distribution"}} and \textit{\textbf{"team coordination"}} during the clinical simulation. Each visualisation provided unique insights that, together, enhanced the understanding of how tasks were managed, prioritised, and coordinated. While the \barchart~ offered an overview of task division and patient prioritisation, the \wardmap~ revealed the spatial dynamics and physical coordination needed to manage tasks effectively.

In one debrief session, the educator used the \barchart~ to explain task distribution and the balance between working together and individually on the main bed: \textit{"Working together on tasks for [the main bed] was quite small, and then working individually on tasks for [the main bed] was a lot bigger,"} highlighting a strategy of task division for efficiency. The \wardmap~ added context by showing how nurses positioned themselves around the main bed and other areas to perform tasks like ECG and observations: \textit{"So, that might be...checking the temperature, checking the pupils, you know, checking the dressing site...doing those assessments."} (SID 20, \allphases).

In phase \phtwo, both visualisations helped educators discuss delegation and re-prioritisation as the secondary nurses joined the team. Initially focused on other patients, nurses had to shift their attention to the main bed: \textit{"So you guys were working on the other patients trying to get a few bits and pieces done. And like }[Red] \textit{said, that you were pulling }[Yellow] \textit{and you told }[Yellow]\textit{ Let's do this. Hang on. Let's do that. Now we're going to go over here }[pointing to the \wardmap~]. \textit{Sometimes you have to re-prioritise... We're reassessing what's going on."} (SID 26).

 In phase \phthree, educators discussed task management and coordination on the main bed. 
The visualisations reflected how individual actions were coordinated toward a common goal. The \barchart~ showed more time spent working individually on the main bed, while the \wardmap~ highlighted increased movement and speech around the bed. An educator explained: \textit{"That doesn't mean that there was only one person at the bedside. That's because [Red] and [Green] were working on individual tasks, but both at the bedside. So we know you were working [on the main bed], but you were doing individual tasks."} (SID 26).
 

\subsection{Educators' Perceived Benefits and Challenges to Adoption}
\label{sec:results-RQ2}

\subsubsection{Perceived benefits in their teaching practice}
\label{sec:results-RQ2-benefits}
Several educators reported that \tool~ enhanced the \textbf{\textit{structure and focus of their debrief sessions}}. For instance, T2 noted that \tool~ \textit{"gave a little bit more structure to how the debrief went,"} suggesting it helped guide the process systematically, focusing on key objectives like task management and coordination. T1 found \tool~ \textit{"pretty accurate showing with the different visualisations, how they did that,"} allowing educators to align feedback with visual data, reinforcing key points.

\tool~ was particularly appreciated for facilitating discussions on \textbf{\textit{specific learning objectives}} like communication and teamwork, which are critical in nursing education. T5 noted \tool's usefulness in discussing these areas: \textit{"the analytics were around communication and teamwork, and they were the main things that we wanted to talk about."} The visualisations made abstract concepts concrete, and T5 mentioned they also helped address misconceptions, such as when \textit{"nurses who ended up caring for the other patients felt quite that they hadn't done anything."} The tool highlighted everyone's contributions, reinforcing the importance of each role in a clinical setting.

\subsubsection{Challenges in adopting \tool }
\label{sec:results-RQ2-challenges}
Three out of five educators initially hesitated \textcolor{black}{to use all features} due to \textbf{\textit{unfamiliarity with technology}}. T1 said, \textit{"I'm not a very technologically savvy person... I did get a bit nervous when I had to do various things,"} while T3 noted, \textit{"technology for me is not, I'm not very good with technology. So it was a little bit of getting used to."} These comments reflect a common barrier among educators unfamiliar with new tools.

However, as they gained experience, educators reported increased comfort and confidence. T1 shared, \textit{"once I got more confident, it got more and more positive and was actually more and more useful,"} and T2 agreed, saying, \textit{"The more I used it, the more comfortable I felt with it."} This highlights that familiarity and practice were key to overcoming initial reluctance, resulting in a more positive experience with \tool.

Another challenge was related to \textcolor{black}{\textbf{\textit{technology limitations}}} in accurately reflecting the data educators wanted to discuss. Two educators felt the visualisations did not always align with their observations or the points they wanted to emphasise. T3 noted, \textit{"Sometimes when it didn't quite represent what we were trying to say... there were times when even the students might have been communicating really well, but they weren't standing near each other."} This highlights a limitation in the devices, occasionally causing a mismatch between observed behaviours and their visual representation.


\subsection{Educators' and Students' Perceived Usefulness, Trust and Accuracy}
\label{sec:results-RQ3}

\subsubsection{Usefulness}
\label{sec:results-RQ3-usefulness}
Educators considered the \textbf{\barchart} useful for summarising team activities, especially task prioritisation and distribution. However, T3 noted that while helpful, the \barchart~ occasionally did not fully align with what was observed during the simulation, indicating some dissonance between the visual data and expected behaviours.
Students also found the chart valuable for reflecting on their team’s prioritisation and actions. Four students mentioned that it provided useful insights into how their team managed tasks and prioritised care.

Educators consistently highlighted the \textbf{\wardmap} as one of the most useful visualisations. It provided a clear spatial representation of team positions and movements, aiding discussions on task management, teamwork, and patient care. T1 noted that the \wardmap~ clearly showed where each team member was during the simulation, helping both educators and students understand spatial dynamics and coordination.
Several students (n=6) appreciated the \wardmap's ability to show positioning and communication patterns. S10 said, \textit{"I really like the level of communication you can see, like the dark and the light spots; it shows me what I was talking the most and which patients I was taking care of."} The \wardmap~ also supported reflection on team dynamics (n=5), as students could see how well they communicated and identify areas for improvement: \textit{"I can reflect on like whereabouts you were and what you could do next time."} (S13).

The \textbf{\SNA} was favoured by educators for its ability to visually represent multidisciplinary communication (nurse-nurse, nurse-doctor, nurse-relative), which is relevant in healthcare settings. T1 noted this visualisation highlighted both good and poor communication patterns, while T2 used it to address misconceptions about student participation. The \SNA~ objectively reflected who spoke, how often, and to whom.

Most students (n=10) appreciated the \SNA~ for identifying areas where communication could improve, helping them reflect on who they communicated with and who they might have neglected. S2 remarked, \textit{"it kind of shows me that maybe I need to like personally try and talk a lot more with my other teammates as well as the patient."} Students also valued seeing which team members were more supportive in communicating their concerns during the scenario. S4 noted, \textit{"you can see how the team plays out. I was the kind of team leader in that scenario."}

The \textbf{\EN}, which highlighted questioning, acknowledging, and escalation behaviours, received mixed reviews. Educators (n=3) used it to discuss communication strategies like closed-loop communication. T2 appreciated its ability to illustrate these strategies but noted challenges in accurately depicting escalation.

Students (n=9) found the \EN~ useful for reflecting on communication behaviours, helping them identify strengths and areas for improvement. They could see how often they engaged in task allocation, questioning, and acknowledging. As S10 noted, \textit{"It clearly shows that task allocation and acknowledging were common, which makes sense because we were always checking in on what needed to be done."} However, some students (n=6) felt the \EN~ lacked context, not showing who was communicating or the specifics of conversations. S12 remarked, \textit{"The visualisation doesn’t capture the full picture — like what were we actually discussing or questioning?"} This led some students (n=4) to prefer the ward map, which provided more specific data. As S5 noted, \textit{"I found the ward map more useful because it showed where everyone was and what they were doing."}

\subsubsection{Accuracy and trust}
\label{sec:results-RQ2-accandtrust}

Concerning the \textbf{\barchart}, educators (n=3) noted that it was generally accurate but that sometimes did not fully capture the complexity of team dynamics or missed specific details. Educators (n=4) trusted the chart when it clearly matched their observations but were cautious when discrepancies arose. Some educators found the \barchart~ less accurate in certain scenarios or when specific timeline sections were selected. Educators (n=3) indicated that while they did not fully distrust the chart, they needed to explain its context to the students due to occasional dissonances with observed behaviours. One educator also pointed out technical limitations, such as the sensors' capabilities or the timing of data capture, which affected their perception of accuracy.

Students (n=10) noted that the \barchart~ accurately reflected their actions during the simulation. These students expressed that the chart aligned well with their memory, particularly in depicting the focus on the deteriorating patient and the differentiation between tasks done individually and collaboratively.
Students generally expressed a moderate to high level of trust. Students (n=11) indicated they trusted the chart as a reliable visualisation for reflecting on their team's performance. 
However, some students (n=6) expressed concerns about the granularity of the information provided. They mentioned that while the chart captured the general prioritisation, they indicated lower trust due to difficulties interpreting the visualisation or because they felt it did not fully account for all the nuanced activities occurring in the simulation.


As for the \textbf{\wardmap}, all five educators found it a reliable representation of team positions and movements, T3 acknowledged occasional minor tracking issues but still found it highly accurate. 

Most students reported that the \wardmap~ accurately represented their communication and positioning during the simulation. S11 noted: \textit{"I think that's extremely accurate. It gets very colourful at the end of [the main] bed, which makes sense."}
They also found a clear alignment between the information illustrated in the ward map and with their memories (\textit{"From what I remember, that's exactly where I was and where I was speaking." - S7; "It looks that is what we did do."} - S14). 
Similar comments were made in terms of trust. Some students (n=6) explained that their high level of trust is particularly aligned with the accuracy of the data and compared to what they remembered from the simulation (\textit{"I would trust it because from what I remember, everything is like how the scenario ran."} - S13). 
Students' sense of trust was also linked with the technology's precision in capturing the location and verbal absence and presence (n=4). They expressed that the sensors were pretty accurate, particularly the tracking sensor (e.g., \textit{"I know it was tracking us and it was listening to when we were talking and stuff, so it seems pretty accurate."} - S14)

The \textbf{\SNA~} was perceived as generally effective in demonstrating communication interactions and patterns. Educators (T3 and T5) noted minor issues, such as missing audio data or timing discrepancies when selecting specific phases or timeline sections, affecting the information's accuracy. 

Most students (n=13) agreed that the visualisation accurately represented their team's communication. They felt the chart correctly depicted their simulation recollections, particularly who they communicated with most frequently and the overall structure of their communication patterns ( \textit{"I was talking to my secondary nurse more about everything."} -S1, \textit{"it aligns with my memory"} - S11).
However, the lack of details \textcolor{black}{in terms of the dialogue content} affected students' accuracy (n=5), expressing their desire for more details (e.g. \textit{"what sort of things we are talking about"} -S7;  \textit{"the quality of communication or information I was giving"} - S12).
Students also stressed the limitations of technology and how it negatively affected the accuracy of the information (e.g., \textit{"I talked to the doctor a lot, but that it didn't show that."} - S4).
A similar trend was observed in terms of trust. Most of the students expressed high trust in the information (n=11), mainly due to the accurate data representation (e.g., \textit{"I completely trust this information because I think it's a really accurate representation of what happened."} - S3)

Finally, while educators found the \textbf{\EN~} useful for discussing specific strategies such as closed-loop communication, their trust was somewhat affected due to the difficulty of interpreting the data. T1 and T3 found difficulties linking the \EN~ to discussions and indicated some uncertainty in interpreting the \EN, but still found them useful as part of a broader discussion. 

Students had mixed perceptions of \EN's accuracy and trust. Some students (n=8) indicated a high accuracy as they noted it accurately represented their team's communication behaviours and aligned with their past experience. However, other students (n=7) indicated moderate accuracy due to lacking depth and details. Students suggested the need for examples (from their behaviour) or explanations on how the data was analysed to improve their accuracy perceptions. 
A similar trend was observed for perceptions of trust. Students (n=8) felt confident that the chart accurately depicted their behaviours and could be used effectively to judge or reflect on their team’s performance. However, other students (n=7) expressed that while the visualisation provided a general understanding of their behaviours, the lack of expertise in interpreting this unfamiliar visualisation affected their overall trust.

\section{Discussion}

\subsection{Summary of Results and Research Questions}
\subsubsection{\textcolor{black}{Supporting educators to facilitate tailored reflective discussions} }
Regarding \textbf{RQ1}, our study found that educators employed a range of strategies with \tool, from simple single-phase debriefing to more complex multi-phase navigation. Most educators preferred guiding discussions using the overall summary (using \textit{all} phases filter) rather than focusing on individual phases. \textcolor{black}{Notably, the \textit{all} phases option is presented by default in the \textit{debrief control view}. One factor contributing to the limited use of multi-phase navigation could be educators' lack of experience using data-driven evidence in debriefing (as discussed later in RQ2). This suggests that, given educators' extensive experience conducting debriefs without additional tools, they were more comfortable relying on the default overall summary rather than using \tool's dynamic filtering (Sec.~\ref{sec:results-RQ1-strategies}). However, as educators became more familiar with \tool
's features, the use of multi-phase strategies also increased.}

Our findings also highlight the effectiveness of various visualisations in supporting evidence-based reflective debriefs. Educators emphasised the importance of communication dynamics and structured communication, often combining \SNA~ and \EN~ visualisations. They used the \SNA~ to analyse multi-disciplinary communication patterns and \EN~ to focus on high-level strategies, such as closed-loop communication (Sections ~\ref{sec:results-RQ1-frequency} and ~\ref{sec:results-RQ1-themes}). This aligns with the literature on teamwork, which emphasises communication as critical for effective patient care \cite{hargestam2013communication,Mazzocco_2009}.
Task management and coordination were also key topics (Sec. ~\ref{sec:results-RQ1-themes}). The combination of the \barchart~ and \wardmap~ visualisations provided a comprehensive view of task distribution and movement within the physical space. As noted by \citet{alonso2012building}, \textit{"communication alone does not constitute teamwork"} (p.42). Effective teams also need to recognise roles and allocate tasks, which educators highlighted during their discussions.

\subsubsection{\textcolor{black}{Perceived benefits and challenges to adoption}}
Regarding \textbf{RQ2}, findings suggest that TeamVision improved the structure and focus of debrief sessions, providing educators with a systematic way to guide discussions on key learning objectives (Sec. ~\ref{sec:results-RQ2-benefits}). This indicates that \tool~ can facilitate the debriefing process, which is crucial given the limited time educators have to prepare narratives and key points to discuss \cite{fraser2018cognitive,Tannenbaum2013}. 

Despite initial concerns about unfamiliarity with the technology, educators' confidence increased with \textcolor{black}{continued use of \tool~ (Sec. ~\ref{sec:results-RQ2-challenges}). \textcolor{black}{This aligns with the observed progression in strategies adopted by educators (Sec. ~\ref{sec:results-RQ1-strategies}), transitioning from single-phase to multi-phase data filtering.} Such behaviour reflects findings from research} on video-based debriefing, where technology is often perceived as a barrier rather than a facilitator of debriefing practices \cite{Abeer2018,KROGH2015180,levett2014systematic}. 
To address this, more training opportunities for nurse educators on AI utilisation and visualisation capabilities, along with effective integration strategies are needed.

A notable challenge was the misalignment between visualisations and observed behaviours. Educators found that sometimes visualisations did not always match their observations, leading to \textcolor{black}{confusion and disruption in their discussions} as they tried to resolve discrepancies \textcolor{black}{(Sec. ~\ref{sec:results-RQ2-challenges})}. This issue, related to the imperfections of AI systems and sensor data \cite{ochoa_multimodal_2022,yan2022scalability,Lee_See_2004}, can disrupt the debrief narrative and risk unfair judgments or misinterpretation of student performance \cite{giannakos2022sensor,Giannakosrole2023}. We explore this issue in the next section.

\subsubsection{\textcolor{black}{Educators' and students' perceptions of usefulness, accuracy and trust.}}
Finally, addressing \textbf{RQ3}, \textcolor{black}{in terms of \textbf{\textit{usefulness}}}, educators and students generally found the \wardmap~ and \SNA~ to be the most \textcolor{black}{useful} visualisations for reflecting on team dynamics and communication, \textcolor{black}{as both depicted individual-level information.} The \EN~ received mixed feedback; while useful for discussing strategies like closed-loop communication, it sometimes lacked context. \textcolor{black}{Although during the design process, educators expressed a preference to avoid reviewing content (Sec.~\ref{sec:design-ENA}), after using \tool, both the educators and students emphasised the importance of detailed content for effective use. These findings suggest the need for a re-design of the \EN~ to include individual-level information or more nuanced content to improve usability and adoption.}
The \barchart~ was found useful but did not always capture the complexity of team dynamics.
\textcolor{black}{During their debriefing use, we found educators combining the \barchart~ with the \wardmap~ to provide more contextual visual information (Sec.~\ref{sec:results-RQ1-frequency}). Therefore, future re-design efforts might consider merging these two visualisations to offer a more comprehensive and insightful representation of team dynamics.}

\textcolor{black}{In terms of \textit{\textbf{accuracy and trust}},} the \wardmap~ was appreciated for its accuracy, aligning closely with recollections of the simulation, while the \SNA~ was seen as reliable for illustrating communication, despite \textcolor{black}{technology} issues (i.e., missing audio data). 
\textcolor{black}{Both educators and students noticed that }technical limitations caused discrepancies between simulation experiences and visualised data, \textcolor{black}{ impacting interpretation of information, accuracy, and trust (Sec.~\ref{sec:results-RQ2-accandtrust}). This finding aligns with the work of \citet{Lee_See_2004}, highlighting how trust in automation systems is shaped by a dynamic interplay between information display and intrinsic trust levels, affecting interpretation and information selection.}
Accuracy and trust are common challenges in AI-powered systems relying on sensor technology as the accuracy of data collection, analysis, and visualisation depends heavily on sensor quality and algorithm reliability \cite{giannakos2022sensor,Lee_See_2004,yangReexaminingWhetherWhy2020} \textcolor{black}{which may affect the adoption of these systems}. If educators and students consider these AI-powered systems as untrustworthy or hard to interpret, their uptake and integration into teaching and learning processes will be limited.
Addressing these challenges requires increasing system transparency and accountability for errors, as suggested in prior research \textcolor{black}{on human-AI interaction} \cite{yangReexaminingWhetherWhy2020,Liao2023}. \textcolor{black}{ Researchers and designers of AI-powered systems in educational contexts could adopt a practical strategy as the one proposed by \citet{Liao_Sundar_2022}, who advocate for designing trustworthiness cues into AI systems for end users. Their strategy focuses on communicating key dimensions of trust, including AI-generated content (e.g., diagnosis accuracy) and transparency (e.g., metrics for accuracy, fairness, and robustness), as integral components for building user trust.}

\subsection{\textcolor{black}{Reflections on the Design Process and In-the-wild Implementation }}
\textcolor{black}{Our study demonstrates how educator involvement in the design process influenced the development of \tool's features, supporting the adoption of various reflection strategies (\textbf{RQ1}) and enabling more structured debrief discussions (\textbf{RQ2}). However, \textit{in-the-wild} use revealed that some features, such as phase filtering, were underutilised despite educators’ active participation in the design process. Multi-phase comparisons were observed only occasionally (\textcolor{black}{Sec.~\ref{sec:results-RQ1-strategies}}). This limited use can be attributed to time constraints \cite{Echeverria2024LAK} and initial unfamiliarity with \tool~ (\textcolor{black}{Sec.~\ref{sec:results-RQ2-challenges})}. These findings highlight that \textbf{\textit{educator-requested features do not always translate into practical use}}, especially in time-constrained, real-world settings. Similarly, while educators indicated during design sessions that they would not require detailed content from linked transcripts, in-the-wild usage changed their minds, as they concluded that this would be useful (Sec.~\ref{sec:results-RQ3-usefulness}). Future design iterations should validate features such as linking transcripts to communication networks, pre-configured templates, short summaries \cite{fernandez2024editor}, and adaptive agents \cite{yan2024vizchat} to better support educators' practices and increase feature adoption.}

\textcolor{black}{In addition, we observed differences in how educators and students perceived the accuracy and trust of visualisations as some discrepancies emerged (\textbf{RQ3}). Educators reported higher trust and accuracy (Sec.~\ref{sec:results-RQ2-accandtrust}), likely due to their involvement in the design process and familiarity with reflective practices. In contrast, students, who were not involved in the design process, required more time to interpret the visualisations and found some data irrelevant due to the lack of details (Sec.~\ref{sec:results-RQ2-accandtrust}). This finding highlights the \textbf{\textit{importance of actively involving both educators and students in the design process}} to ensure the system supports their needs, thereby improving trust, usability, and adoption \cite{Alfredo2024HCLA_review}.}

\subsection{Implications for Supporting Reflection Practices}

There is so much that HCI research can offer to learning analytics research, especially via in-the-wild studies \cite{BuckinghamShum2024hci}. Our study highlights key implications and opportunities for \textit{\textbf{healthcare education}}. AI-powered systems, as shown by our findings, can enhance educators' practices by aiding in event recollection and supporting debrief discussions. Advances in AI offer new ways to improve teaching and learning in medical simulations \cite{Benfatah2024,Jagannath2022}, allowing educators to focus on specific team dynamics and provide tailored feedback. Unlike traditional video-based debriefs, which require reviewing entire footage and pausing for discussions \cite{Abeer2018,KROGH2015180,levett2014systematic}, \tool~ facilitates more efficient navigation through critical moments and summarised information, potentially reducing learner discomfort associated with video-assisted tools.



\textcolor{black}{\tool~ opens new opportunities for supporting reflection in \textbf{\textit{broader classroom contexts}}, where communication and movement are key aspects of teaching and learning practice. \tool~ could be adapted to team teaching, a collaborative instructional approach where multiple educators share responsibilities within the same physical classroom \citep{Dang2022, steeleWhatMakesCoTeaching2021}. By analysing communication patterns, task coordination, and spatial interactions during classroom activities, \tool~ can provide educators with evidence to reflect on their co-teaching strategies, enhancing both collaborative practices and teaching outcomes.
Unlike other tools that focus only on communication \citep{Ngoon2024classInsight} or spatial data \citep{Alfredo2025TeamTeachingViz}, \tool~ integrates both modalities to enable more comprehensive evidence-based reflection. However, effective implementation will require adapting algorithms to capture key communication strategies unique to this context and addressing challenges related to educator training, data privacy, and the seamless integration of \tool~ into existing reflection practices without increasing workload.}


\subsection{Implications for HCI research}
This study contributes to CSCW and HCI research, which has long focused on how digital technologies support effective teamwork \cite{carroll2005teamwork,Fitzpatrick2013cscw}. Specifically, it adds to recent HCI work on using AI to enhance teamwork practices \cite{Khakurel2022augmenting}. Building on the HC-AI design process outlined in Section \ref{Design}, our research explores the deployment of AI-powered and multimodal learning analytics (MMLA) systems in real-world educational settings.

Our work represents one of the most extensive deployments of MMLA/AI systems, both within and beyond healthcare education, as noted by recent reviews \cite{yan2022scalability,Schneider2024advances}. This study goes beyond previous frameworks \cite{Huceta2022} that provided conceptual foundations for human-centred teamwork analytics, by demonstrating the practical feasibility of such AI-powered systems.
Our findings show that \tool~ effectively enhances reflective practices and enables more structured, focused debriefing sessions. This addresses the need for evidence on teamwork in healthcare simulation, quantitative performance measures, and scalable debriefing tools \cite{Moslehi_2022,cant2017valueOfSim, hegland2017simSLR,hildreth2023telEmergency,Martin_2020}, and responds to calls for novel HCI techniques in medical training \cite{Sadeghi2022hcihealthcare}.

The system’s scalability suggests potential applications in various team-based learning environments beyond healthcare simulations. However, future research should address challenges related to the accuracy and trustworthiness of AI models, particularly with sensor-based data prone to errors. Ensuring system reliability is crucial for broader adoption and responsible integration into educational settings \cite{Shneiderman_2020,yangReexaminingWhetherWhy2020}. Future HCI research must also consider the ethical and practical implications of integrating AI-driven systems while balancing human expertise and automation.

\subsection{Limitations and Future Work}
Our study has several limitations. In terms of generalisability, \tool~ was designed specifically for a healthcare simulation, which may limit its applicability to other domains. While it could be adapted for other team-based environments, modifications would be needed. Additionally, the design focused on fostering non-technical skills such as teamwork and communication. Further adaptations \textcolor{black}{to the dialogue content analysis or visualisations} would be required if the focus shifted to technical skills or clinical decision-making, presenting opportunities for future research to explore \tool's use in diverse educational contexts.

Moreover, the impact of \tool~ in our study is limited by its short-term, real-world application. This study does not capture the extended influence of the AI-powered system on teaching and reflection practices or long-term outcomes as learners were exposed to only a single high-fidelity simulation. Longitudinal studies and controlled experiments are needed to understand \tool's sustained contribution to skill development in real-world healthcare settings. 
Such experiments could explore quantitative measures under different experimental conditions. Despite these limitations, the \textit{in-the-wild} study demonstrated \tool's practical applicability and potential. 

\textcolor{black}{Finally, \tool's technical limitations should be acknowledged, as they may influence the interpretation of data and affect the usefulness, accuracy and trustworthy of information, potentially hindering broader adoption. As discussed in Section 7.1, transparency mechanisms to communicate and build users' trust are needed in these AI-powered educational systems. }


\section{Concluding Remarks}
This study provides evidence on \tool, an AI-powered system, designed in collaboration with educators. \tool supports data-driven discussions during reflective practices in team-based healthcare simulations. Educators found \tool~ valuable for structuring debriefings and providing evidence on team's behaviours. Both educators and learners generally found the visualisations from \tool useful and trustworthy, though those \textcolor{black}{which presented group-level information} were seen as less accurate, indicating a need for details and contextual information. This research highlights the potential for AI-powered analytics to enrich reflective practices and lays the groundwork for future innovations aimed at improving the learning experience in complex, team-based environments.

\bibliographystyle{ACM-Reference-Format}
\bibliography{sample-base}

\end{document}